\documentclass[a4paper,12pt]{article} 
\pdfoutput=1

\usepackage{pdfpages}
\usepackage{graphicx}
\usepackage{amsmath, amsthm, amssymb}
\usepackage{setspace}  
\usepackage[super]{natbib}
\bibpunct{}{}{,}{a}{,}{,}

\usepackage{epsfig}

\newtheorem{Prop}{Proposition}
     
\newcommand{\be}{\begin{equation}}
\newcommand{\ee}{\end{equation}} 
\newcommand{\lb}{\label}
\newcommand{\ol}{\overline}

\newcommand{\btau}{{\mbox{\boldmath $\tau$}}}

\newcommand{\const}{({\rm const.})}

\newcommand{\bk}{{\bf k}}

\newcommand{\br}{{\bf r}}
\newcommand{\bu}{{\bf u}}
\newcommand{\bw}{{\bf w}}

\newcommand{\bv}{{\bf v}}
\newcommand{\bx}{{\bf x}}
\newcommand{\by}{{\bf y}}

\newcommand{\bp}{{\bf p}}
\newcommand{\bq}{{\bf q}}

\newcommand{\grad}{{\mbox{\boldmath $\nabla$}}}
\newcommand{\bdot}{{\mbox{\boldmath $\cdot$}}}

\newcommand{\bell}{{\mbox{\boldmath $\ell$}}}

\begin{document}

%\maketitle                  

%\tableofcontents

%\doublespacing

\baselineskip=18pt
\renewcommand{\thefootnote}{\alph{footnote})}

\begin{center}
\begin{Large}
{\bf Localness of energy cascade in hydrodynamic turbulence, II.
Sharp spectral filter} \\
\end{Large}

\bigskip
\bigskip

Hussein Aluie\footnote{Email: hussein@jhu.edu}
and Gregory L. Eyink\footnote{Email: eyink@ams.jhu.edu} \\
{\it
Department of Applied Mathematics \& Statistics\\ 
The Johns Hopkins University \\
3400 North Charles Street \\
Baltimore, MD 21218-2682\\
}

\bigskip
\bigskip

\begin{abstract}
We investigate the scale-locality of subgrid-scale (SGS) energy flux 
\textcolor{black}{and inter-band energy transfers} 
defined by the sharp 
spectral filter.  We show by rigorous bounds, physical arguments and numerical simulations
that the spectral SGS flux is dominated by local triadic interactions in an extended turbulent 
inertial-range. 
\textcolor{black}{Inter-band energy transfers are also shown to be dominated by local triads if the 
spectral bands have constant width on a logarithmic scale.} 
We disprove in particular an alternative picture of  ``local transfer by nonlocal triads,'' 
with the advecting wavenumber mode at the energy peak. Although such triads have the largest 
transfer rates of all {\it individual} wavenumber triads, we show rigorously that, due to their
restricted number, they make an asymptotically negligible contribution to energy flux 
\textcolor{black}{and log-banded energy transfers} 
at high wavenumbers in the inertial-range. We show that it is only the aggregate effect of a 
geometrically increasing number of local wavenumber triads which can sustain an energy 
\textcolor{black}{cascade}
to small scales. Furthermore, non-local triads are argued to contribute even less to the 
space-average energy flux than is implied by our rigorous bounds, because of additional 
cancellations from scale-decorrelation effects. We can thus recover the -4/3 scaling of 
nonlocal contributions to spectral energy flux predicted by Kraichnan's ALHDIA and TFM closures. 
We support our results with numerical data from a $512^3$ pseudospectral simulation 
of isotropic turbulence with phase-shift dealiasing.
We also discuss a rigorous counterexample of Eyink (1994), which showed that non-local 
wavenumber triads may dominate in the sharp spectral flux (but not in the SGS energy 
flux for graded filters). We show that this mathematical counterexample fails to satisfy
reasonable physical requirements for a turbulent velocity field, which are employed in 
our proof of scale-locality. We conclude that the sharp spectral filter has a firm theoretical
basis for use in large-eddy simulation (LES) modeling of turbulent flows. 
\end{abstract}
\end{center}

\vspace{0.5cm}

%\bigskip\noindent

\clearpage

\clearpage

\newpage

%\doublespacing

\section{Introduction}\lb{intro}

This paper is the second part of a study of the scale-locality properties of the 
energy cascade of three-dimensional turbulence. In the first part \cite{EyinkAluie} (hereafter 
referred to as I) we investigated energy transfer defined by means of low-pass and 
band-pass filters for smooth, dilated kernels, which permit simultaneous resolution 
of the physical processes both in space and in scale.   We demonstrated there the 
scale-locality of interactions involved in turbulent energy transfer, by rigorous 
analysis and by numerical simulation. See also \cite{Eyink95a,Eyink05}. 
The discussion in these papers did not apply to sharp-spectral filters, which 
do not satisfy the modest decay conditions in physical space that were employed 
there. The origin of scale-locality with the sharp spectral filter is, in fact,  a rather  
more subtle issue.  The problem is of importance, however, since most numerical 
studies of turbulent scale-locality employ sharp spectral filters. We believe that 
misinterpretation of the numerical results has led to a number of 
misunderstandings in the literature. 

Early papers on turbulent energy cascade which employed spectral analysis---such as 
those of Obukhov \cite{Obukhov41a,Obukhov41b}, Onsager \cite{Onsager45,Onsager49}, 
and Heisenberg \cite{Heisenberg48}---argued for scale-locality. Those works proposed a 
``cascade process'' in which two modes of similar wavenumber transfer energy to a mode 
with twice that wavenumber, implying steps in spectral space of increasing size with increasing 
wavenumber. This picture was later confirmed by detailed closure calculations of Kraichnan, 
both for his (Eulerian) direct-interaction approximation (DIA) \cite{Kraichnan59},  his 
abridged Lagrangian-history direct-interaction approximation (ALHDIA) \cite{Kraichnan66}, 
and the test-field model (TFM) \cite{Kraichnan71a}. In the late 80's and early 90's, 
however, this traditional view was contested in numerical simulation studies by a number
of authors: Brasseur \& Corrsin \cite{BrasseurCorrsin}, Domaradzki \& Rogallo
\cite{DomaradzkiRogallo}, Yeung \& Brasseur \cite{YeungBrasseur}, and Ohkitani \& Kida
\cite{OhkitaniKida}. Those works presented results suggesting that energy cascade is dominated 
by highly nonlocal triadic interactions, with one very energetic mode at the lowest wavenumber 
$k_0$ catalyzing transfer between two modes at high wavenumbers $k<k',$ for $k'-k=O(k_0).$
This transfer process was described as  ``local transfer through nonlocal interactions''. 
In contrast to the traditional picture, the energy flow through spectral space was suggested 
to proceed via steps of a fixed, small size $k_0,$ driven by low-wavenumber advection. 
Chasnov \cite{Chasnov91} argued that such nonlocal, advective sweeping interactions 
should be incorporated into spectral large-eddy simulation (LES) via a stochastic force 
that produces random backscatter of energy. If such views were  correct, they would 
present a fundamental challenge to the Kolmogorov picture of local energy cascade 
and the universality of small-scale turbulence. 

Several theoretical papers published shortly thereafter defended the traditional 
view of turbulent scale-locality  \cite{LvovFalkovich,Waleffe92,Zhou93a,Zhou93b,Eyink94}.
In particular, Waleffe \cite{Waleffe92}, Zhou \cite{Zhou93a,Zhou93b} and Eyink \cite{Eyink94}
all argued that scale-locality is recovered when Fourier modes are suitably combined 
or averaged.  Zhou \cite{Zhou93a,Zhou93b} showed numerically that locality does 
hold,  when some necessary summations are made, and he verified the quantitative 
prediction of Kraichnan \cite{Kraichnan66,Kraichnan71} that non-local contributions
to energy flux decay as a power $\propto s^{4/3}$ of the ``scale-disparity parameter'' 
$s=k_{\min}/k_{{\rm med}}$  (where $k_{\max},k_{{\rm med}},k_{\min}$ are the maximum, 
median, and minimum wavenumbers of a triad, respectively). It seems, however, that 
this conclusion was  subjected to doubt in a subsequent investigation by Zhou et al. 
(1996) \cite{Zhouetal96} which indicated  that the effects of the largest scales are 
significant. The paper of Eyink (1994)\cite{Eyink94} proved rigorously that spectral 
energy flux, if  averaged in wavenumber over an octave band, is dominated by local 
triads. His argument yielded a rigorous upper bound $O(s^{2/3})$ on the nonlocal
triadic contributions, which is larger than Kraichnan's scaling prediction but still vanishing
as $s\rightarrow 0$. However, the paper of Eyink also proved that the spectral energy flux, 
without additional averaging, may be non-local.  He constructed a velocity field with H\"{o}lder 
continuity properties analogous to those observed in turbulent flow for which the 
instantaneous spectral energy flux is dominated by nonlocal advective interactions, 
somewhat similar to the effects seen in the numerical simulations 
\cite{BrasseurCorrsin,DomaradzkiRogallo,YeungBrasseur,OhkitaniKida}.

The debate about the locality of interactions contributing to energy flux with the sharp
spectral filter has  continued unabated. Recently, the issue was addressed by numerical
simulations at much higher Reynolds numbers by Alexakis et al. (2005) \cite{Alexakisetal05b} 
and Mininni et al. (2006,2008) \cite{Mininni06, Mininni08}, who concluded, in agreement with 
the earlier simulations \cite{BrasseurCorrsin,DomaradzkiRogallo,YeungBrasseur,OhkitaniKida},
that the important interactions 
\textcolor{black}{contributing to interband transfers} 
are those among highly non-local wavenumber triads 
with one leg at the energy peak. \textcolor{black}{On the other hand,  they concluded 
that the spectral energy flux, involving  a suitable summation of  Fourier modes, is 
dominated by local triads.} 
Domaradzki \& Carati (2007,2009) 
\cite{Domaradzki07b, Domaradzki09} \textcolor{black}{confirmed these results and furthermore
numerically calculated} Kraichnan's \cite{Kraichnan66, Kraichnan71} locality 
function $W(s),$  a quantity which measures the fractional contribution of  nonlocal triads. 
They verified Kraichnan's prediction that $W(s)\propto s^{4/3}$ for $s\ll 1,$ 
in agreement with earlier studies of Zhou \cite{Zhou93a,Zhou93b}. On the 
other hand, they found no difference between the sharp spectral filter and various graded 
filters, in apparent contradiction with the analytical work of Eyink\cite{Eyink94,Eyink95,Eyink05} 
who drew a clear distinction between the sharp spectral filter and smooth, dilated filters. 

It may be an understatement to say that no coherent picture of the scale-locality 
of turbulent energy cascade is presented by the current literature on the subject. Open issues
that continue to be discussed include: the need for additional averaging (over space,time, 
ensembles,etc.), qualitative distinctions between sharp spectral versus smoothly graded
filters, the quantitative fraction of energy flux contributed by non-local triads, and possible 
differences between spectral binning with logarithmic versus linear bands. Even the proper 
definition of  ``energy flux'' is unclear, since the definition used in the rigorous proofs with 
graded filters \cite{Eyink95,Eyink05,EyinkAluie} is the ``SGS flux'' familiar from large-eddy 
simulation modeling, which involves additional advective subtractions to define the 
subscale stress which are not  present in the conventional definition of the spectral 
energy flux. 

The present paper has two main purposes.
Our first goal is to present new results on scale-locality of spectral 
\textcolor{black}{transfer quantities}, including 
rigorous estimates, physical arguments and numerical simulation data.  In particular, we 
shall show rigorously that \textcolor{black}{both energy flux and inter-band transfers} 
defined by the sharp spectral filter---without any additional averaging---\textcolor{black}{are}
dominated by local triadic interactions.  Our proof of locality 
rests on four ingredients: (1) The SGS flux (defined in I and in the next section) as the unique Galilean 
invariant measure of the inter-scale energy transfer at each point in the flow, (2) scaling properties
that are observed empirically to hold for the turbulent velocity field, (3) wavenumber conservation, 
which constrains the number of Fourier modes contributing to the energy flux, and (4) the 
essential use of ``logarithmic'' spectral bands whose width increases proportional to wavenumber. 
The second aim of this paper is to attempt to collect the disparate results in the literature 
into a clear and consistent picture of the local cascade process in three-dimensional turbulence.
We shall try to provide answers to many of the open issues mentioned above which continue
to be debated in the literature. 

To aid this latter goal, it may be useful to summarize here the essential reason for the 
dominance of local triadic interactions in \textcolor{black}{energy cascade}. It is  true, 
as indicated by numerical studies  \cite{BrasseurCorrsin,DomaradzkiRogallo,YeungBrasseur,
OhkitaniKida,Alexakisetal05b,Mininni06, Mininni08}, that \emph{individual} non-local triads make 
a much bigger contribution to energy flux than \emph{individual} local ones,  due simply to their
having one of their modes at the energy containing scales. However, such comparisons of single 
triads vastly underestimate the aggregate contribution from local triads. Whereas the number 
of highly elongated, nonlocal triads  grows moderately with increasing wavenumber, the number 
of local triads increases much more rapidly. The net contribution to energy flux from the exploding 
population of local triads dwarfs the contribution from the smaller number of nonlocal triads. While 
this paper was being written, we discovered a work of Verma et al.\cite{Vermaetal} which arrives at 
essentially the same conclusion, by means of a non-rigorous perturbative closure analysis of energy 
transfer.  They stated there that  ``...the shell-to-shell energy transfer rate  is found to be local and forward. This result is due to the fact that the nonlocal triads occupy much less Fourier space volume
 than the local ones.'' \textcolor{black}{A related observation was also made by Alexakis et al.\cite{Alexakisetal05b}, who observed that the fractional contribution of non-local triads to energy 
 flux is reduced ``since many more local triads contribute in the global summation.''}
In fact, these ideas, as we shall discuss, are already implicit in the arguments advanced 
by Kraichnan for scale-locality, using his ALHDIA and TFM closures\cite{Kraichnan66,Kraichnan71}. 
Our work supports these conclusions and extends them, by rigorously bounding the nonlocal contributions to spectral flux, without any heuristic approximations. \textcolor{black}{Our analysis
implies as well the dominance of local triads to inter-band energy transfers, for wavenumber 
bands of constant width on a logarithmic scale.} We also explain by physical arguments how 
decorrelation effects (as discussed in I for the graded filter) further diminish the nonlocal 
contributions, giving agreement with Kraichnan's asymptotic scaling predictions. 

We verify our  theoretical analysis by analyzing the velocity field $\bu$ generated 
from a direct numerical simulation of the incompressible Navier Stokes equation
\be  \partial_{t} \bu +  (\bu\cdot\grad)\bu = -\grad p + \nu \nabla^2 \bu + {\bf f} 
\lb{NS-eq} \ee
$$\grad\cdot\bu=0$$
which is solved pseudo-spectrally in a periodic box of $512^3$ grid-points. Here, 
$p$ is the pressure, $\nu$ is the viscosity, and ${\bf f}$ is an external stirring force.
We advance in time using the second-order Adam-Bashforth scheme and employ
the phase-shift method to remove aliasing effects \cite{PattersonOrszag}. The fluid is 
stirred using Taylor-Green forcing: 
\be{\bf f} \equiv f_0 [ \sin(k_fx)\cos(k_f y)\cos(k_f z) {\hat \bx} - \cos(k_fx)\sin(k_fy)\cos(k_fz) {\hat \by}]
\lb{forcing}\ee
where $f_0 = 2$ is the force amplitude, and $k_f = 2$. The viscosity $\nu$ is 
$0.87\times10^{-3}$ and the Reynolds number $Re\equiv u_{rms}L/\nu \approx 11800$,
where $L=2\pi$ is the size of the simulation box and $u_{rms}=\sqrt{2E_{tot}}=1.63$ is the rms velocity. 
The Reynold's number based on the Taylor scale $\lambda = 2\pi  \sqrt{E_{tot}}/ \big(\int dk k^2 E(k)\big)^{1/2}$
is $Re_\lambda = 615$.
We analyze a time snapshot of the flow after it has reached a statistically steady state. 
The energy spectrum $E(k)$ of the flow, shown in Figure \ref{FigureSpectrum}, has a reasonable 
$k^{-5/3}$ scaling until around $k=40$, before dissipation effects start to dominate.
The mean energy flux $\ol{\Pi}_K=\langle\Pi_K\rangle$ is shown in Figure \ref{FigureFlux}. 

\begin{figure}
\begin{center}
\epsfig{figure= 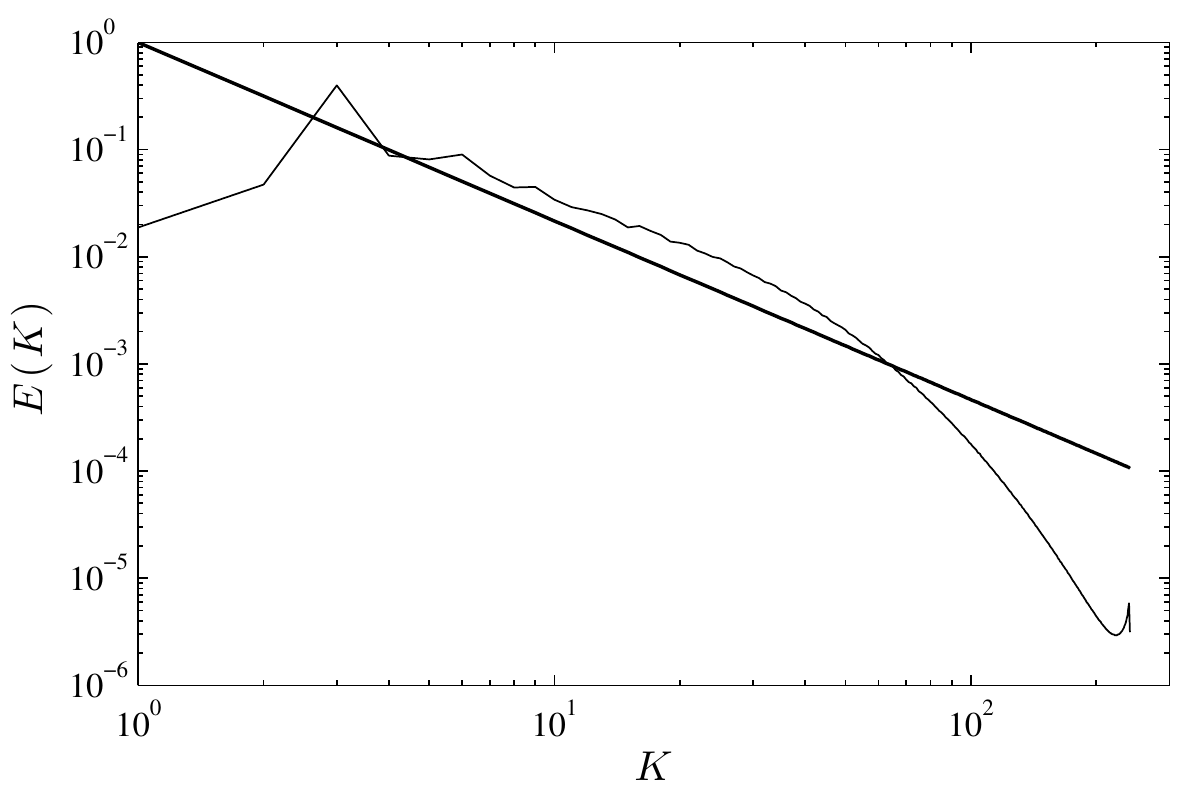}%,width=400pt,height=400pt } %, clip="[92 345 938 742]"}
\end{center}
\caption{ Energy spectrum for the simulation being analyzed. The forcing is at $k_f= 2$. 
A straight line with slope $-5/3$ is added for reference.
}
\lb{FigureSpectrum}
\end{figure}

\begin{figure}
\begin{center}
\epsfig{figure= 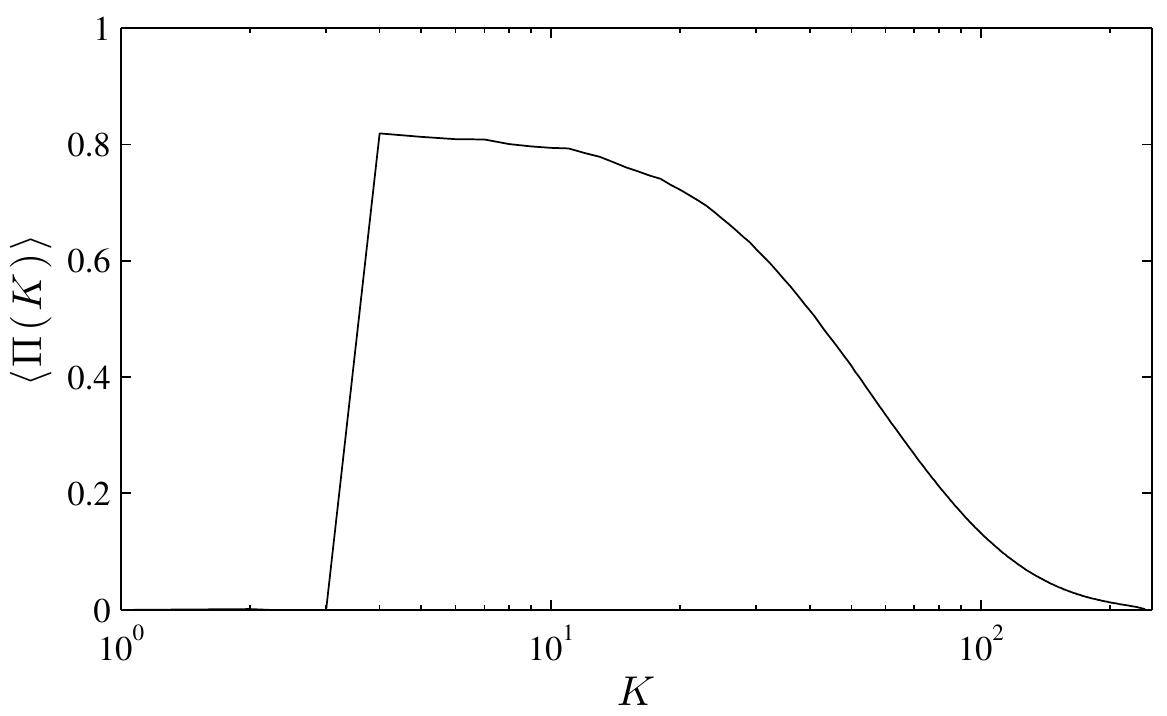}%,width=400pt,height=400pt } %, clip="[92 345 938 742]"}
\end{center}
\caption{ Average energy flux $\langle\Pi_K\rangle$ from the simulation. }
\lb{FigureFlux}
\end{figure}

The outline of our paper is as follows. Our main new results are presented in the following Section II. 
Subsection II.A contains some preliminary definitions and discussion. In subsection II.B we establish 
ultraviolet locality of spectral SGS flux, and in subsection II.C we establish infrared locality. In the next 
section III, we examine how our work relates to previous research on scale-locality, in particular the 
analysis of Kraichnan\cite{Kraichnan66} using spectral closure and recent numerical studies\cite{Alexakisetal05b,
Mininni06, Domaradzki07a,Domaradzki07b, Mininni08,Domaradzki09}. Section IV formulates our final 
conclusions.Two appendices contain some technical material, Appendix A on our rigorous proofs and 
Appendix B on Kraichnan's ALHDIA closure.

\section{Main Analytical and Numerical Results}\lb{results}
%\section{Scale locality using the sharp spectral filter}

\subsection{Preliminaries}

In a periodic domain ${\Bbb T}^3=[0,2\pi)^3$, one may represent a velocity field $\bu$ by  
its Fourier series expansion:
$$ \bu(\bx)=\sum_{\bk\in {\Bbb Z}^3} \widehat{\bu}(\bk) e^{i\bk\bdot\bx}. $$
Restricting this sum to wavenumbers satisfying $|\bk|\leq K$ yields
the {\it low-pass filtered field} $\bu^{<K}$, while restricting the sum to $|\bk|>K$ 
gives the complementary {\it high-pass filtered field} $\bu^{>K} = \bu -\bu^{<K}$.
We shall also employ in our discussion {\it band-pass filtered fields} 
$$ \bu^{[K,Q]}=\bu^{>K}-\bu^{>Q}=\bu^{<Q}-\bu^{<K} $$
for $K<Q.$ We employ a special notation $[K]$ for dyadic (octave) bands $[K/2,K]$ 
and denote the corresponding band-pass velocity field by $\bu^{[K]}.$  The reader 
may assume here that $|\bk|$ denotes the Euclidean norm $|\bk|=
\sqrt{k_x^2+k_y^2+k_z^2}$. In fact, due to very subtle aspects of harmonic 
analysis, our proofs do not apply with full rigor for the Euclidean norm, although 
they are rigorous for the norms $|\bk|_1=|k_x|+|k_y|+|k_z|$ or $|\bk|_\infty=
\max\{|k_x|,|k_y|,|k_z|\}.$ See Appendix A for a careful discussion of these delicate 
mathematical issues. In fact, we believe that our arguments do apply with the Euclidean 
norm on wavenumbers and that the abstract mathematical counterexamples that 
compromise their rigor have different properties from those enjoyed by the turbulent
velocity fields. In support of this belief, all of the numerical results presented later 
shall employ the Euclidean norm and, as may be seen, they agree with our theoretical 
analysis. 

Just as with any other filtering scheme, one may employ the sharp spectral filter
to decompose the Navier-Stokes equation (\ref{NS-eq}) into large-scale and 
small-scale components and then consider the corresponding energy balances. 
Following a standard analysis, presented in I,  one arrives at the sub-grid scale 
(SGS) energy flux 
\be \Pi_K(\bu,\bv,\bw) \equiv -\partial_i u_j^{<K} \bigg( (v_i w_j)^{<K} 
         -v_i^{<K} w_j^{<K} \bigg) \lb{sgs} \ee
as the representation of the transfer rate of energy from modes with wavenumbers $<K$ to 
modes with wavenumbers $>K$ at each space point. We have distinguished here a 
{\it straining mode} $\bu,$ an {\it advecting mode} $\bv,$ and an {\it advected mode} $\bw,$ 
although, in the physical balance equation, $\bu=\bv=\bw.$ This permits us to formulate 
the property of scale-locality in a precise manner. As in I, we say that energy flux is 
{\it ultraviolet (UV) scale-local} if replacing any single $\bu$ or any combination of $\bu$'s in 
$\Pi_K(\bu,\bu,\bu)$ by $\bu^{[P]}$ gives an asymptotically negligible contribution 
for $P\gg K.$ Likewise, we say that energy flux is {\it infrared (IR) scale-local} if replacing 
any single $\bu$ or any combination of $\bu$'s in $\Pi_K(\bu,\bu,\bu)$ by $\bu^{[Q]}$ 
gives an asymptotically negligible contribution for $Q\ll K.$ 

Most of the previous studies of scale-locality with the sharp spectral filter have employed 
what we call an ``unsubtracted flux'' 
$$    \Pi^{uns}_K = -(\partial_i u_j)^{<K} (u_i u_j) \,\,\,\,\, {\rm or} \,\,\,\,\,
        -(\partial_i u_j)^{<K} (u_i u_j)^{<K} . $$ 
These two quantities have the same average over space as the SGS energy flux, because the 
sweeping terms which distinguish them integrate to zero: $\int d\bx~ \partial_i u_j^{<K}  u_i^{<K} 
u_j^{<K} = 0$. However, the pointwise values are different and the locality properties are 
different as well. Note that the quantities $\Pi_K(\bu,\bu,\bu^{[Q]})$ and $\Pi^{uns}_K 
(\bu,\bu,\bu^{[Q]})= -\partial_i u_j^{<K} \big(u_i u^{[Q]}_j \big)^{<K} $ are not the same 
even after space-averaging, since the sweeping effects no longer integrate to zero:
$\int d\bx~ \partial_i u_j^{<K}  u_i^{<K}(u^{[Q]}_j)^{<K}  \neq 0$, as was recognized 
by Domaradzki and Carati \cite{Domaradzki07a}. Indeed, it is easy to see that the quantity 
$\Pi^{uns}_K$ cannot be scale-local instantaneously, at each point in the space domain. 
If we boost the flow with a uniform velocity $\bu_0$ (a $k=0$ mode), then $\Pi^{uns}_K$ 
will develop a contribution directly proportional to this velocity. Since $\bu_0$ can be made 
arbitrarily large, triads involving the $k=0$ mode can give the dominant contribution at any 
fixed $K$. This argument does not apply to the space (or ensemble) average of  $\Pi^{uns}_K$,
which {\it is} Galilei invariant. As we shall discuss later, the mean value $\ol{\Pi}^{uns}_K$
is in fact scale-local, as concluded by early workers \cite{Obukhov41a,Obukhov41b,Onsager49,
Heisenberg48,Kraichnan59,Kraichnan66}, although its locality properties are less 
robust than those of the SGS flux $\Pi_K.$ We examine only the latter for all of our 
detailed estimates in this section, and we return to discuss $\Pi^{uns}_K$ in section 
\ref{discuss}. 
 
Scale-locality of interactions is not a general property \textcolor{black}{of solutions} of the 
Navier-Stokes equation, e.g. the nonlinear interactions in laminar flow are dominated 
by the most energetic modes. Rather, the locality of \textcolor{black}{the energy cascade} 
in a turbulent flow depends crucially on \textcolor{black}{some scaling properties of the 
solution.} The operative scaling laws for sharp-spectral filtered quantities are closely 
related to those for velocity increments  \textcolor{black}{$\delta \bu(\bell;\bx)=
\bu(\bx+\bell)-\bu(\bx).$} Heuristically,
$$\bu^{>K} \simeq \bu^{[K]}\simeq \delta u(\ell_K), \,\,\,\,\,\,\, \grad \bu^{<K} \simeq 
     \delta u(\ell_K)/\ell_K $$ 
with  $\ell_K=2\pi/K$ the length-scale corresponding to wavenumber $K.$ More precisely,
the scaling laws that are used in our arguments below are for volume-averages:
\be \langle \big| \bu^{>K} \big |^{p} \rangle^{1/p}, 
     \langle \big| \bu^{[K]} \big |^{p} \rangle^{1/p} \sim ({\rm const.}) u_{rms} (KL)^{-\sigma_p}, \,\,\,\,
    \langle \big| \grad \bu^{<K} \big |^{p} \rangle^{1/p}  \sim ({\rm const.}) 
    \frac{u_{rms}}{L} (KL)^{\textcolor{black}{1-\sigma_p}}
\lb{scaling} \ee
with $0<\sigma_p<1$ and $p\geq 1,$ for inertial-range wavenumbers $K$ satisfying 
$KL\gg 1.$ Here $L$ is the integral length-scale,  $u_{rms}$ is the root-mean-square 
velocity, and $\langle . \rangle$ is a volume average. Note that $\sigma_p = \zeta_p/p,$  where $\zeta_p$ is the scaling 
exponent of the $p$th-order absolute velocity structure function:  
$\langle [\delta u(\ell)]^p \rangle \sim \ell^{\zeta_p}$. The ultimate source of these scaling 
properties is empirical evidence from experiments and numerical simulations \cite{Anselmetetal,Chenetal,Sreenivasanetal}. The precise mathematical forms of the scaling 
laws (\ref{scaling}) that are used in our proofs are discussed in Appendix A. Here 
we emphasize just one salient point: the scaling law stated for $\langle \big| \bu^{[K]} \big |^{p} 
\rangle^{1/p} $ in (\ref{scaling}) is valid only for  logarithmic bands $[K]$ and does  not hold for 
a band $[K,K+\Delta]$ with fixed spectral width $\Delta.$ This point is essential for the proof 
of infrared locality, as we shall see.

\subsection{Ultraviolet Locality}

We shall begin by treating the case of ultraviolet locality, which is somewhat easier and 
allows us to introduce the basic method of argument. 

\subsubsection{Theoretical Analysis}

We must  consider the quantities $\Pi_K(\bu^{[P]},\bu,\bu),\Pi_K(\bu,\bu^{[P]},\bu),$ and 
$\Pi_K(\bu,\bu,\bu^{[P]})$ for $P\gg K$ to establish UV locality in the straining, advecting, 
and advected modes, respectively. The first of these is trivial, since $\partial_i(u_j ^{[P]})^{<K} =0$ 
for $P > 2K$ and thus such modes give a vanishing contribution to the strain. The second two UV 
locality properties are less trivial but their physical origin lies in the simple fact that modes 
at smaller scales contain less energy. Thus, the contribution from modes 
$P\gg K$ to the stress at wavenumber $K$
$$\btau_K(\bu,\bu)=(\bu\bu)^{<K}-\bu^{<K}\bu^{<K} $$ 
is small. We will now show this through precise estimates.

We first consider $\Pi_K(\bu,\bu,\bu^{[P]})=-\partial_i u_j^{<K} \tau_K(u_i,u_j^{[P]}).$ 
When $P > 2K$ the subtraction in the stress vanishes, so that there is no difference 
between the SGS flux and the ``unsubtracted'' flux. This quantity then reduces to:
$$  \Pi_K(\bu,\bu,\bu^{[P]}) =-\partial_i u_j^{<K}    \big(u_i u_j^{[P]} \big)^{<K} 
=  -\partial_i u_j^{<K}    \big(u_i^{[\frac{P}{2}-K,P+K]} u_j^{[P]}\big)^{<K}. $$
The condition that the advecting mode lie in the band $[P/2-K,P+K]$ arises from 
the constraint that the wavenumbers of the two velocity modes in the stress must 
sum to a value $<K.$ We thus see that, for $P\gg K,$ the two stress 
modes must lie in essentially the same high-wavenumber band, since 
$[P/2-K,P+K]$ is the same as $[P]=[P/2,P]$ with a little extra padding of size $K$. 
This argument applies with equal force to the quantity 
$$  \Pi_K(\bu,\bu^{[P]},\bu) =-\partial_i u_j^{<K}    \big(u_i^{[P]} u_j\big)^{<K} 
=  -\partial_i u_j^{<K}    \big(u_i^{[P]} u_j^{[\frac{P}{2}-K,P+K]} \big)^{<K}. $$
Thus, we may consider these two quantities together.  

A rigorous estimate which establishes UV locality of the SGS flux follows from the 
H\"{o}lder inequality: 
$$ \big\langle |\partial_i u_j^{<K}    \big(u_i^{[\frac{P}{2}-K,P+K]}u_j^{[P]}\big)^{<K} |
\big\rangle \le ({\rm const.}) ~\big\langle |\nabla \bu^{<K} |^3\big\rangle^{1/3} 
 \big\langle |\bu^{[\frac{P}{2}-K,P+K]}|^3 \big\rangle^{1/3}\big\langle |\bu^{[P]}|^3
\big\rangle^{1/3}. $$
See Appendix A for details. Together with the scaling laws (\ref{scaling}) this gives
\be  \big\langle |\Pi_K(\bu,\bu,\bu^{[P]})|\big\rangle 
\le \const \frac{u_{rms}}{L}(LK)^{1-\sigma_3} ~  u_{rms}^2 (LP)^{-2\sigma_3} 
~\sim  \varepsilon \const (LK)^{1-3\sigma_3} ~ (K/P)^{2\sigma_3}. \lb{UV-local} \ee
where $\varepsilon \equiv u_{rms}^{3}/L$ is the energy injection rate.  For any scaling
exponent $\sigma_3>0$, the factor $(K/P)^{2\sigma_3}$ becomes very small when 
$P\gg K,$ implying that the high $P$ modes make asymptotically little contribution to the 
space-averaged energy flux.  In fact, it is known empirically that $\sigma_3\doteq 1/3,$ 
the K41 value. Thus, we obtain a bound close to $O((K/P)^{2/3}).$

One difficulty in the above estimation is the factor $(LK)^{1-3\sigma_3}.$ Since 
$\sigma_3\lesssim 1/3,$ this factor grows slowly with increasing $K$ and has the potential 
to render our bound useless at very high $K,$ deep in the inertial range. This would 
be the case if the bound were larger than the total energy flux $\varepsilon$ itself, so that 
it would cease to provide a tight constraint on the large-$P$ contribution. Since the growth 
in $K$ is so slow, however, it can be easily compensated by taking $P$ large enough. More 
precisely, define the wavenumber $P_*(K)= K(KL)^{(1-3\sigma_3)/2\sigma_3}\geq K.$
Then our upper bound (\ref{UV-local}) is $\ll \varepsilon$ for $P\gg P_*(K),$ implying 
that such modes make an asymptotically negligible  contribution to the mean flux. This 
proves rigorously the UV-locality of the space-average energy flux. Note, however, that 
the bound (\ref{UV-local}) is probably far from optimal, as discussed more below. 

The same argument as above can be applied to prove that 
$$ \langle |\Pi_K(\bu,\bu,\bu^{[P]})|^{p} \rangle^{1/p} 
      \leq \varepsilon \const (LK)^{1-3\sigma_{3p}} ~ (K/P)^{2\sigma_{3p}} $$ 
for any $p\ge1$ (Appendix A). In the limit $p\to \infty$, the quantity 
$\langle |\Pi_K(\bu,\bu,\bu^{[P]})|^{p} \rangle^{1/p} $ increases to 
$\sup_{\bx}|\Pi_K(\bu,\bu,\bu^{[P]})|,$ the maximum over the domain,  
and $\sigma_p$ decreases to $h_{\min},$ the minimum H\"{o}lder exponent 
of the velocity field. Therefore, by taking larger $p$, our rigorous estimates hold 
more uniformly over space but also imply less rapid decay with increasing $P.$
These results are similar to those established by Eyink\cite{Eyink95,Eyink05} and
in I for graded filters, but weaker. In particular, we do not have pointwise estimates
of energy flux at each space point $\bx$ in terms of the local H\"{o}lder exponent, 
using spectral filters. 

Up until this point, our discussion has been mathematically rigorous. It is often the 
case, however, that rigorous proofs do not yield the optimal results on complex  
physical problems. Our upper bound $O((K/P)^{2\sigma_3})$ on the mean flux 
contributed by high-wavenumber modes $P\gg K$ is larger than the asymptotic 
scaling result $\propto (K/P)^{4/3}$ predicted by Kraichnan\cite{Kraichnan66,Kraichnan71}.
Just as discussed in I for graded filters, we can argue on heuristic, physical grounds  that 
the smaller contribution found by Kraichnan is due to cancellations that arise in the 
average over space, which are neglected in our crude upper bound. 
\textcolor{black}{For details, see I and \cite{AluieThesis}.}
\textcolor{black}{The result is that
$$  \langle \Pi_K(\bu,\bu,\bu^{[P]}\rangle\sim 
                          \varepsilon \const \bigg(\frac{\ell_P}{\ell_K}\bigg)^{\zeta_1 +1}. $$}
%To derive the last expression, we have used the fact that  $\zeta_3 \doteq 1$, which is 
%known from empirical measurements \cite{Anselmetetal,Chenetal,Sreenivasanetal}.
If we assume also the K41 value $\zeta_1=\frac{1}{3},$ then we recover the 4/3 scaling prediction  
of Kraichnan\cite{Kraichnan66,Kraichnan71} who obtained both the fractional 
contribution $(\ell_P/\ell_K)^{2\zeta_2}$ and the K41 value $\zeta_2=\frac{2}{3}$ 
from his ALHDIA and TFM closures. Both of these results also agree with the 
$(\ell_P/\ell_K)^{2-\zeta_2}$ estimate made by  L'vov and Falkovich\cite{LvovFalkovich}, 
if one assumes $\zeta_2=\frac{2}{3}$. In principle, however, all of these results are slightly 
different due to intermittency effects and experimentally distinguishable.

\subsubsection{Numerical Results}

We shall now present results from our $512^3$ DNS, both to investigate
the sharpness of our rigorous bounds and to test the validity of the physical 
arguments. 

We first present simulation data for the following quantities: 
\be \big|\partial_i u_j^{<K}    \big(|u_i^{[\frac{P}{2}-K,P+K]}u_j^{[P]} |\big)^{<K} \big|, \,\,\,\,\,
      \big|\partial_i u_j^{<K}    \big(|u_i^{[P]} u_j^{[\frac{P}{2}-K,P+K]}| \big)^{<K} \big|. \lb{UV-abs} \ee
All of our rigorous estimates apply to these two objects, but neither can experience the additional
cancellations invoked in our heuristic argument, because of the added absolute values.
Notice that we have included absolute values even inside the low-pass filter $<K.$
As discussed in paper I, substantial cancellations could otherwise occur in the space-integral 
which defines that low-pass filter.  We have calculated space-averages of the quantities (\ref{UV-abs}) 
for $K=4$ and for logarithmic bands $[P]=[P/2,P]$ varied continuously over values from 
$P=8$ to $P=240$, where $k_{max}=241$ is the maximum wavenumber in the simulated flow field.
We plot the results in Figures \ref{FigureUV_point}a, b, respectively, normalized by 
$\langle |\Pi_K|\rangle$. The two quantities are very similar and both show a slightly faster
decay than our prediction. This is consistent with the findings of Domaradzki \& Carati 
\cite{Domaradzki07b,Domaradzki09}, who also see faster decay of UV nonlocal contributions
than predicted by Kraichnan\cite{Kraichnan66,Kraichnan71}. This is plausibly attributed 
to the extreme shortness of the inertial range and contamination by viscous effects. The 
decay rate beyond $P=40$ is definitely increased by viscosity. We have fitted power-laws
over the range $P=8$ to $P=40$, which is the putative ``inertial-range'' of our DNS 
based on the scaling of the energy spectrum (Figure \ref{FigureSpectrum}). Over this 
range we obtain a $P^{-0.83}$ scaling for the first quantity in (\ref{UV-abs}) and $P^{-0.98}$ 
for the second, both distinctly faster than $P^{-2/3}.$

\begin{figure}
\begin{center}
\epsfig{figure= 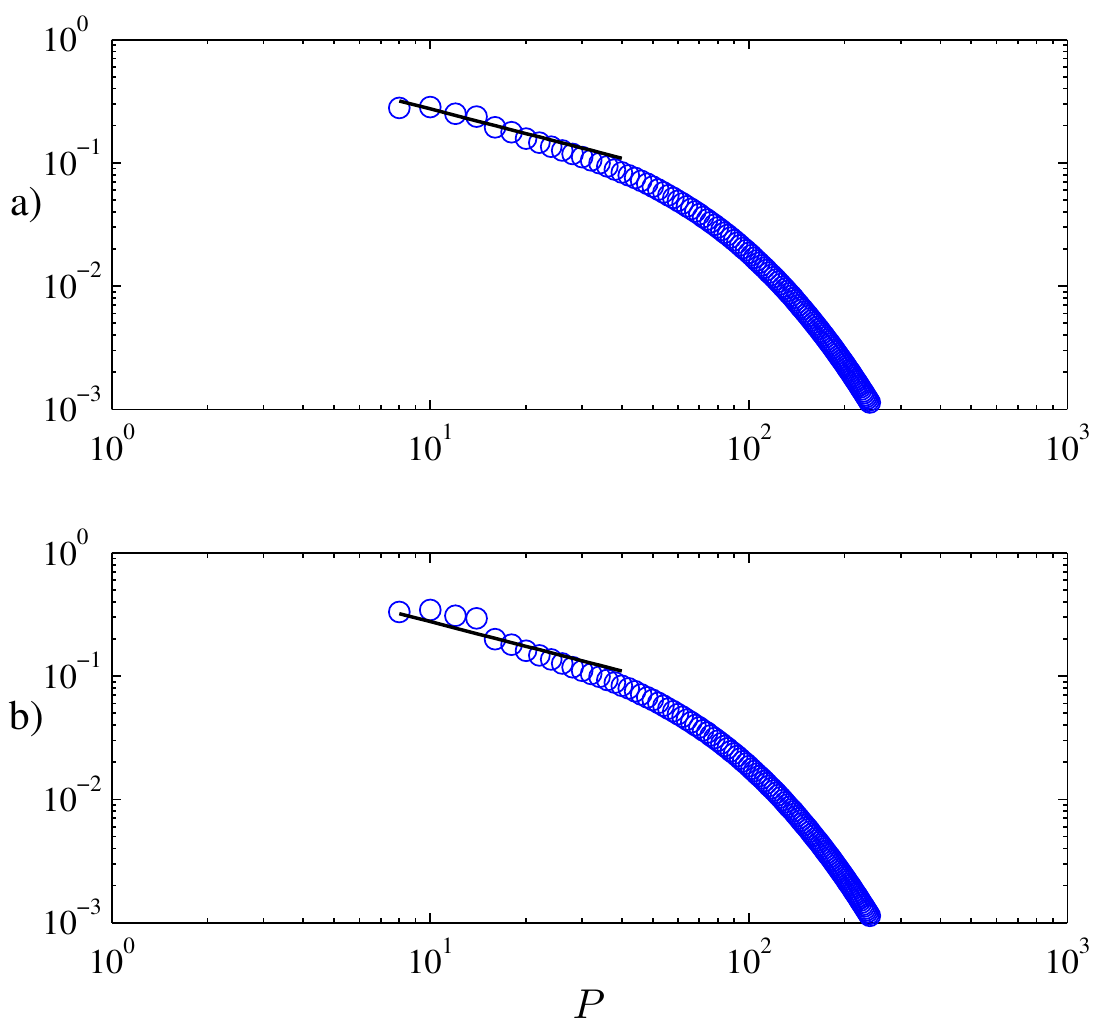}%,width=400pt,height=400pt } %, clip="[92 345 938 742]"}
\end{center}
\caption{ For $K=4$, we plot  (a) $\langle \big|\partial_i u_j^{<K} \big(|u_i^{[\frac{P}{2}-K,P+K]}u_j^{[P]}|\big)^{<K} 
\big| \rangle/\langle|\Pi_K|\rangle$  and (b) $\langle\big|\partial_i u_j^{<K}  \big(|u_i^{[P]}u_j^{[\frac{P}{2}-K,P+K]}| 
\big)^{<K} \big|\rangle/\langle|\Pi_K|\rangle$ using logarithmic bands $[P/2,P]$. The straight lines with $-2/3$ 
slope are for reference  and extend over the fitting region which gives  slopes of -0.83 and -0.98 for (a) and (b),
respectively. The two plots are essentially identical for large $P$,  as expected.
}
\lb{FigureUV_point}
\end{figure}

We next present results for the mean quantities $\ol{\Pi}_K (\bu,\bu,\bu^{[P]})$ and 
$\ol{\Pi}_K (\bu,\bu^{[P]},\bu),$ again for $K=4$ and for $P=8 - 240,$ normalized by the 
mean flux $\ol{\Pi}_K=\varepsilon.$ These data are plotted in Figure \ref{FigureUV} 
and show good agreement with our predictions. When fitted over the ``inertial range'' 
from $P=8$ to $P=40$,  we obtain a $P^{-1.54}$ scaling for  $\ol{\Pi}_K (\bu,\bu,\bu^{[P]})$ 
and $P^{-1.42}$ for $\ol{\Pi}_K (\bu,\bu^{[P]},\bu).$  This is a slightly faster decay 
than $P^{-4/3}$, again in accord with the recent results by Domaradzki \& Carati 
\cite{Domaradzki07b,Domaradzki09}.  Note that the plots for $\ol{\Pi}_K (\bu,\bu,\bu^{[P]})$ 
and $\ol{\Pi}_K (\bu,\bu^{[P]},\bu)$ are identical for large $P$.  This makes good sense, because, 
as mentioned above, the maximum and middle wavenumbers can differ at most by $K$ 
and thus those two modes arise from nearly the same wavenumber range.  Another 
interesting feature is that for $P\ge 110$, the partial flux becomes negative. This shows up 
as a kink on the log-log plot, a feature that was also observed by Domaradzki \& Carati  
\cite{Domaradzki09}.

\begin{figure}
\begin{center}
\epsfig{figure= 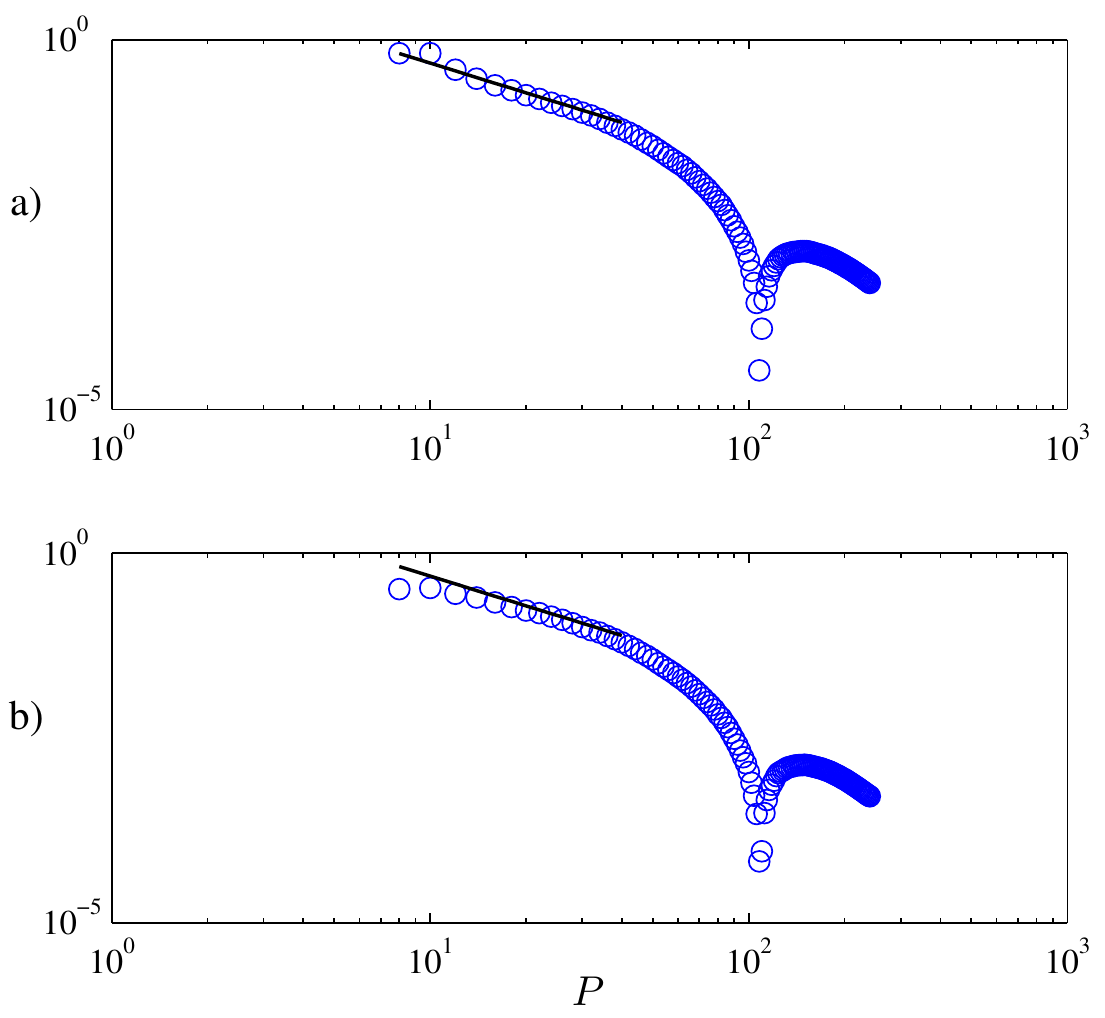}%,width=400pt,height=400pt } %, clip="[92 345 938 742]"}
\end{center}
\caption{ For $K=4$, we plot the global quantities (a) $\ol{\Pi}_K (\bu,\bu,\bu^{[P]})/\ol{\Pi}_K$ 
and (b) $\ol{\Pi}_K (\bu,\bu^{[P]},\bu)/\ol{\Pi}_K$ using logarithmic bands $[\frac{P}{2},P]$. 
The straight lines with $-4/3$ slope are for reference  and extend over the fitting region, which 
yields slopes of -1.54 and -1.42 for (a) and (b) respectively. The two plots are essentially identical 
for large $P$. The kink at $P\simeq 110$ in the log-log plot is due to a change in sign at that wavenumber.
}
\lb{FigureUV}
\end{figure}

\subsection{Infrared Locality}

We turn now to a discussion of the infrared locality of the spectral energy flux. We shall prove,
by similar arguments as in the previous section, that the large-scale components of all three velocity 
modes---the straining, advecting and advected modes---contribute negligibly to spectral 
\textcolor{black}{energy flux}.  
The importance of the large-scale advecting mode contribution has, in particular, been a major source 
of contention in the literature. Several studies  \cite{BrasseurCorrsin,DomaradzkiRogallo,YeungBrasseur,
OhkitaniKida,Alexakisetal05b, DomaradzkiRogallo, OhkitaniKida, Mininni06} employing numerical simulations 
have concluded that nonlocal advective interactions are primarily responsible for the energy transfer. 
In this picture, energy cascades to smaller scales by very many small steps in wavenumber, of fixed size, 
mediated by the largest, energy containing modes. If true, such a picture would have dramatic implications 
for Kolmogorov's concept of small-scale universality and on the theory and practice of LES modeling.
However, we \textcolor{black}{argue} below (and in section \ref{discuss}) that this picture is false, and 
is \textcolor{black}{an artefact of the use of constant (unit) width spectral bands to define 
energy flux and energy transfer.}

\subsubsection{Theoretical Analysis}

We derive exact bounds on the quantities $\Pi_K(\bu^{[Q]},\bu,\bu)$, $\Pi_K(\bu,\bu^{[Q]},\bu)$, 
and $\Pi_K(\bu,\bu,\bu^{[Q]}) $ for $Q\ll K$, thus establishing IR locality in each velocity mode.

We first consider the contribution of the straining mode:
\be    \Pi_K(\bu^{[Q]},\bu,\bu)  =   -\partial_i u_j^{[Q]} \tau_K(u_i,u_j). \ee
The origin of IR-locality here is fairly obvious, because the strain from low-wavenumbers is weak. 
This fact may be expressed as a rigorous upper bound by using the H\"{o}lder inequality: 
$$ \bigg\langle \big| \partial_j u_i^{[Q]} \tau_K(u_i,u_j) \big|^p \bigg\rangle^{1/p}
 \le \bigg\langle \big| \grad\bu^{[Q]}\big|^r \bigg\rangle^{1/r}  \bigg\langle \big| 
 \tau_K(\bu,\bu) \big|^s \bigg\rangle^{1/s} $$
with $\frac{1}{p}=\frac{1}{r}+\frac{1}{s}$. Because $ \tau_K(u_i,u_j) \sim \delta u_i(\ell_K)\delta u_j(\ell_K),$ 
it is expected that 
\be \bigg\langle \big| \tau_K(u,u) \big|^s \bigg\rangle^{1/s}\sim \const u_{rms}^2 (KL)^{-\rho_s} 
\lb{stress} \ee
with $\rho_s\doteq 2\sigma_{2s}.$ See Meneveau \&  O'Neil\cite{MeneveauONeil94} and our own  
figure \ref{FigureStress}, where we verify this scaling for a sharp spectral filter with $s=2.$ 
If this result is combined with the scaling relation (\ref{scaling}) for the velocity-gradient, we obtain
\begin{eqnarray}
\bigg\langle \big| \partial_j u_i^{[Q]} \tau_K(u_i,u_j) \big|^p \bigg\rangle^{1/p} 
     & \le &  \const  \frac{u_{rms}}{L}(LQ)^{1-\sigma_r} ~  u_{rms}^2(KL)^{-\rho_s} \cr
     &  =  & \const \varepsilon (Q/K)^{1-\sigma_r} (KL)^{1-\sigma_r-\rho_s}. 
\lb{strain-UV} \end{eqnarray}      
As long as $\sigma_r<1,$ the factor $(Q/K)^{1-\sigma_r}$ is decaying for $Q\ll K,$ implying 
IR locality in the straining mode.  Just as in the discussion of UV locality, larger $p$ corresponds 
to estimates more uniform in space. In the IR case, however, the decay estimates {\it improve} 
for increasing $p$ because $\sigma_p$ is non-increasing in $p$ (Frisch \cite{Frisch}, section 8.4).
For example, with $r=s=2p,$ we see that the exponent $1-\sigma_{2p}$ 
increases with $p$ and the bound on the nonlocal contribution becomes tighter. 

Also as in the UV case, there is an additional overall factor $(KL)^{1-\sigma_r-\rho_s}$ 
which causes our bound to deteriorate for $KL\gg 1.$ For example, for $p=1,r=s=2,$
$\sigma_2+\rho_2\doteq \sigma_2+2\sigma_4$ and the latter is slightly less than $1$
by the concavity of $\zeta_p$ as a function of $p.$ (See, for example, Frisch \cite{Frisch}, 
section 8.4). However, our bound (\ref{strain-UV}) is still useful (less than $\varepsilon$) 
if we take $Q$ small enough to offset the slight growth in $K$. Defining the wavenumber 
$Q_*(K)=(1/L)(KL)^{2\sigma_4/(1-\sigma_2)}$ such that $1/L\ll Q_*(K)\leq K$, one need only take 
$Q\ll Q_*(K)$ to ensure that the contribution of straining modes at wavenumber $Q$ make
a negligible contribution to energy flux. 

We now discuss the large-scale contributions of the other two velocity modes which appear
through the turbulent stress, i.e. the advecting and advected modes. These two cases are 
far more delicate. We shall consider in detail the case of the advecting mode, which has been 
the source of most of the controversy in the literature. Thus, we shall examine 
\be    \Pi_K(\bu,\bu^{[Q]},\bu)  =   
-\partial_i u_j^{<K}  \bigg[ \big(u_i^{[Q]}u_j\big)^{<K}  - (u_i^{[Q]})^{<K}u_j^{<K} \bigg] 
\lb{cnvctng} \ee
for $Q<K/2.$ The case of the advected mode is quite similar and will be discussed in 
parallel. The key to the whole analysis is the following fundamental identity: 
\be  \Pi_K(\bu,\bu^{[Q]},\bu) 
=-\partial_i u_j^{<K}  \bigg[\big(u_i^{[Q]}u_j^{[K-Q,K+Q]}\big)^{<K}  - u_i^{[Q]}u_j^{[K-Q,K]}\bigg]. 
\lb{fund-id} \ee
To derive this relation, note that any advected modes $u_j$ with wavenumbers $<K-Q$ 
in (\ref{cnvctng}) when multiplied by $u_i^{[Q]}$ will contain only wavenumbers $<K.$ 
Hence, the low-pass filter $<K$ on the first term in the square bracket $[\cdot]$ can 
be dropped and the contribution of all such modes cancels exactly with the similar 
contribution from the second term in the bracket. The additional restriction to 
wavenumbers $<K+Q$ for $u_j$ in the first term arises from the fact that modes with 
wavenumbers $>K+Q$ when multiplied by $u_i^{[Q]}$ can only give wavenumbers 
$>K$ and these modes do not survive the low-pass filter $<K.$ The identity (\ref{fund-id})
expresses a basic restriction on the {\it range} of interactions allowed to very
nonlocal triads, with advecting modes at low wavenumber $Q$ interacting only with
advected modes in a band of wavenumbers $[K-Q,K+Q]$ at most, with width 
$2Q.$ A similar identity can be derived as well for  $\Pi_K(\bu,\bu,\bu^{[Q]})$,  by the 
same argument. This restriction on the range of the nonlocal triadic interactions is the 
origin of the IR locality of the SGS spectral flux in the advecting and advected modes. 

We now wish to demonstrate this by a rigorous estimate. Unfortunately, the scaling 
relations (\ref{scaling}) that we have employed until now are not applicable to estimate
the size of the band-pass field $\bu^{[K-Q,K+Q]}.$ As we mentioned earlier and discuss
in more detail in Appendix A, the scaling of $u^{[K]}$ in (\ref{scaling}) follows from the 
empirically-known scaling of velocity-structure functions, but only for  $[K]=[K/2,K],$ a logarithmic 
band. Here,  the band $[K-Q,K+Q]$ has a fixed width $2Q$, independent of $K$.  
However, we may appeal to another empirical observation on the scaling of the 
turbulent energy spectrum. A great many experiments and simulations have shown that 
\be E(k,t) \sim C u_{rms}^2 L\,(kL)^{-(1+\zeta_2)} \lb{spectrum} \ee
with $\zeta_2\doteq 2/3,$ for wavenumber binning of {\it unit} size. That is, the Kolmogorov
power-law spectrum is observed not just with octave bands but with bands of essentially
infinitesimal width.   This is, in fact, one of the most  robust empirical observations on 
small-scale turbulence. From it we infer that for $Q\ll K$
$$ \big\langle |\bu^{[K-Q,K+Q]}|^2\big\rangle=2\int_{K-Q}^{K+Q} dk\, E(k,t) 
       \sim \const u_{rms}^2 (Q/K) (KL)^{-2\sigma_2}. $$
The important point here is the factor $(Q/K)$ which becomes smaller with decreasing 
$Q.$ To exploit this estimate, we use the 4-4-2 H\"{o}lder inequality  to derive the bound 
$$ \bigg\langle \big|  \Pi_K(\bu,\bu^{[Q]},\bu)  \big| \bigg\rangle
 \le  \const ~\bigg\langle \big| \grad\bu^{<K}\big|^4 \bigg\rangle^\frac{1}{4}  
      \bigg\langle \big| \bu^{[Q]} \big|^4  \bigg\rangle^\frac{1}{4}
        \bigg\langle \big|\bu^{[K-Q,K+Q]} \big|^2 \bigg\rangle^\frac{1}{2}  $$
$$\sim \const \frac{u_{rms}}{L}(LK)^{1-\sigma_4} \cdot u_{rms} (LQ)^{-\sigma_4}
      \cdot u_{rms} (Q/K)^{1/2} (KL)^{-\sigma_2} $$
\be \sim \const \varepsilon (LK)^{1-\sigma_2-2\sigma_4}  (Q/K)^{1/2-\sigma_4}. 
     \,\,\,\,\,\,\,\,\,\,\,\,\,\,\,\,\,\,\,\,\,\,\,\,\,\,\,\,\,\,\,\,\,\,\,\,\,\,\,\,\,\,\,\,\,\,\,\,\,\,\, \lb{main} \ee
See Appendix A for more details. Since $\sigma_4$ is known empirically to be a 
bit smaller than $1/3,$ the exponent $1/2-\sigma_4>1/6$ and thus the factor 
$(Q/K)^{1/2-\sigma_4}$ becomes small for $Q\ll K.$ This proves the IR locality 
in the advecting mode of spectral SGS flux. The result is due to two competing 
factors. On the one hand, the low-wavenumber velocity modes have greater 
magnitude, reflected in the factor $u_{rms} (LQ)^{-\sigma_4}$ which grows for
decreasing $Q.$ On the other hand, the modes with low wavenumber $Q$ are 
restricted to interact with a smaller set of modes at wavenumber $K,$ reflected 
in the decreasing factor $(Q/K)^{1/2}$. The latter factor wins the competition, 
implying IR locality. This is one of the most important results established in our paper. 

As with our earlier bounds, there is a troublesome factor $(LK)^{1-\sigma_2-2\sigma_4}$
that grows with $K.$ However, defining $Q_*(K)\equiv K(KL)^{2\sigma_2/(1-2\sigma_4) -2}$ 
so that $1/L\ll Q_*(K)\leq K,$ we see that our bound becomes $\ll \varepsilon$ for $Q\ll Q_*(K).$
Thus, IR locality follows, although the result is presumably not optimal. Eyink\cite{Eyink05} 
(and I) used graded filters to derived a sharper bound of the form $O((Q/K)^{2/3}),$ whereas
the decay proved here for the sharp spectral filter is only like $(Q/K)^{1/6}.$ As we shall 
discuss below, there is some evidence from our numerics to suggest that the 2/3 decay rate 
holds also for absolute spectral flux and even faster decay is expected for mean spectral
flux due to decorrelation effects.   

First, however, let us discuss briefly how the bound (\ref{main}) is reconciled with the 
1994 counterexample of Eyink\cite{Eyink94} which showed that spectral flux may be dominated
by nonlocal advective sweeping interactions.  The counterexample cannot satisfy the bound 
(\ref{main}), which would imply instead that the local triads dominate. There is no contradiction, 
however. Eyink's counterexample is a Fourier-Weierstrass-type function given by a lacunary Fourier 
series, with just two wavenumber modes in each octave band $[2^N,2^{N+1}]$ for $N=1,2,3,...$. 
It has scaling properties similar to those of a turbulent velocity field (like (\ref{scaling})), but it fails to 
satisfy the strong condition (\ref{spectrum}) on the energy spectrum because its wavenumber modes 
do not densely populate Fourier space. In this respect, the counterexample appears very different 
from typical turbulent velocity fields. Thus, it is true as a general mathematical fact that energy 
flux defined by the sharp spectral filter may be dominated by nonlocal triads (unlike flux defined 
with graded filters). However, this is not the case for the turbulent fields satisfying condition
(\ref{spectrum}). In particular, the counterexample of Eyink is only slightly related to the ``local 
transfer by nonlocal triads'' observed in turbulent simulations\cite{BrasseurCorrsin,DomaradzkiRogallo,
YeungBrasseur,OhkitaniKida,Alexakisetal05b, DomaradzkiRogallo, OhkitaniKida, Mininni06}.  
As we shall discuss in section \ref{discuss}, those DNS observations are completely compatible 
with locality of energy cascade. 

All of our results on IR locality have, to this point, been mathematically rigorous upper bounds. 
\textcolor{black}{However, we expect much faster decay of non-local contributions for the 
space-average flux than the bounds proved above, due to cancellation of fluctuating positive 
and negative parts. It is possible to provide more physical arguments to explain what we believe 
is the true scaling of IR non-local contributions to mean spectral flux. For details, see I and 
\cite{AluieThesis}. Here we note just the final results that}
% It is 
% worthwhile, however, to provide more physical arguments to explain what we believe is the 
% true scaling of IR non-local contributions to mean spectral flux, just as we did earlier for 
% UV non-local contributions. 
% We expect much faster decay of non-local contributions for the space-average flux than the 
% bounds proved above, due to cancellation of fluctuating positive and negative parts.
%
% We begin with a discussion of IR locality in the strain mode.  We can decompose this partial flux 
% into two separate terms: 
% \be   \ol{\Pi}_K(\bu^{[Q]},\bu,\bu) =   \ol{\Pi}_K(\bu^{[Q]},\bu,\bu^{>K}) +  
% \ol{\Pi}_K(\bu^{[Q]},\bu,\bu^{<K}) 
%\lb{mean-IR-strain} \ee
\textcolor{black}
{\be \ol{\Pi}_K(\bu^{[Q]},\bu,\bu), \,\,\,\,\ol{\Pi}_K(\bu,\bu^{[Q]},\bu) \sim
\varepsilon  \bigg(\frac{\ell_K}{\ell_Q}\bigg)^{1+\zeta_1}.  \lb{mean-IR} \ee}
\textcolor{black}{
The result is different for locality in the advected mode, so we give a few details. By wavenumber 
conservation with $Q<K$:
$$ \ol{\Pi}_K(\bu,\bu,\bu^{[Q]})
      = \langle \partial_i u_j^{[Q]} u_i^{>K}u_j ^{[K-Q,K]} \rangle
       =\langle\partial_i u_j ^{[Q]}  u_i^{>K}u_j^{>K-Q}\rangle  
       - \langle\partial_i u_j ^{[Q]}  u_i^{>K}u_j^{>K}\rangle. $$
The limited number of modes contributing to $u_j^{[K-Q,K]}$  leads to a major cancellation.
Indeed, expressing the triple correlation heuristically as increments and using fusion rules
as in I, we obtain the scaling result for $Q\ll K$}
\begin{eqnarray*}
\ol{\Pi}_K(\bu,\bu,\bu^{[Q]}) &\sim & u^3_{rms} \frac{\ell_K}{\ell^2_Q} 
\bigg(\frac{\ell_Q}{L}\bigg)^{\zeta_3} 
\Big[    \bigg(\frac{\ell_{K-Q}}{\ell_Q}\bigg)^{\zeta_1} - 
\bigg(\frac{\ell_{K}}{\ell_Q}\bigg)^{\zeta_1} \Big] \cr
 &\sim& \varepsilon \frac{\ell_K}{\ell_Q}  \Big[  \frac{\ell_K}{\ell_Q}\cdot 
 \bigg(\frac{\ell_{K}}{\ell_Q}\bigg)^{\zeta_1}   \Big] 
 =  \varepsilon  \bigg(\frac{\ell_K}{\ell_Q}\bigg)^{2+\zeta_1 }. 
\end{eqnarray*} 
\textcolor{black}{The cancellation leads to an additional factor of $\ell_K/\ell_Q,$ which 
gives a faster decay for this non-local contribution.} 

It is worth noting that a similar cancellation
appears in the partial flux $\ol{\Pi}_K(\bu,\bu^{[Q]},\bu),$ where it is crucial to give the 
scaling $\sim (\ell_K/\ell_Q)^{1+\zeta_1}$ in (\ref{mean-IR}).
The physics behind IR locality in the advecting mode 
is  that when $Q$ is very small, then, due to wavenumber conservation, only a few wavenumber 
modes within distance $Q$ of $K$ contribute to the flux, implying insignificant transfer from  
those triads. In section \ref{discuss} we shall rederive this result within the ALHDIA closure 
of Kraichnan\cite{Kraichnan66}, which embodies in a quantitative form the same basic ideas 
employed heuristically above. As for the UV case, our prediction of the IR non-locality correction 
scaling as $(Q/K)^{1+\zeta_1}$  can in principle be distinguished from the scaling prediction 
$(Q/K)^{2-\zeta_2}$ of Kraichnan\cite{Kraichnan66} and L'vov \& Falkovich\cite{LvovFalkovich},
due to small effects of intermittency.

\subsubsection{Numerical Results}

We now present simulation results on IR locality. 
We first consider space-averages of the absolute values of the partial fluxes. These should 
be devoid of any cancellation effects but are constrained by our rigorous upper bounds.  
In Fig.\ref{FigureIR_point} we plot $\langle \big| \Pi_K(\bu^{[Q]},\bu,\bu) \big| \rangle$ 
($\circ$), $\langle| \Pi_K(\bu,\bu,\bu^{[Q]}) |\rangle$ ($\times$) and 
$\langle| \Pi_K(\bu,\bu^{[Q]},\bu)|\rangle$ ($+$), for $K=100$ and for $Q$ ranging 
continuously over all wavenumbers $Q\leq K/2$.  
Fitting with power-laws over the ``inertial-range'' from $Q=8$ to $Q=40$ gives scalings 
of $Q^{0.77}$, $Q^{0.42}$ and  $Q^{0.45},$ respectively. The first is in good agreement 
with our rigorous estimate $(Q/K)^{1-\sigma_2},$ for $\sigma_2\doteq 1/3$. 
The other two quantities have scaling exponents exactly between the value $2/3$ 
and the $1/6$ exponent of our rigorous upper bound.  In Fig.\ref{FigureStress} we 
also test the scaling prediction (\ref{stress}) of the spectral SGS stress for $s=2$ and 
obtain good agreement with the K41 value exponent $\rho_2\doteq 2\sigma_4\doteq 2/3.$

\begin{figure}
\begin{center}
\epsfig{figure= 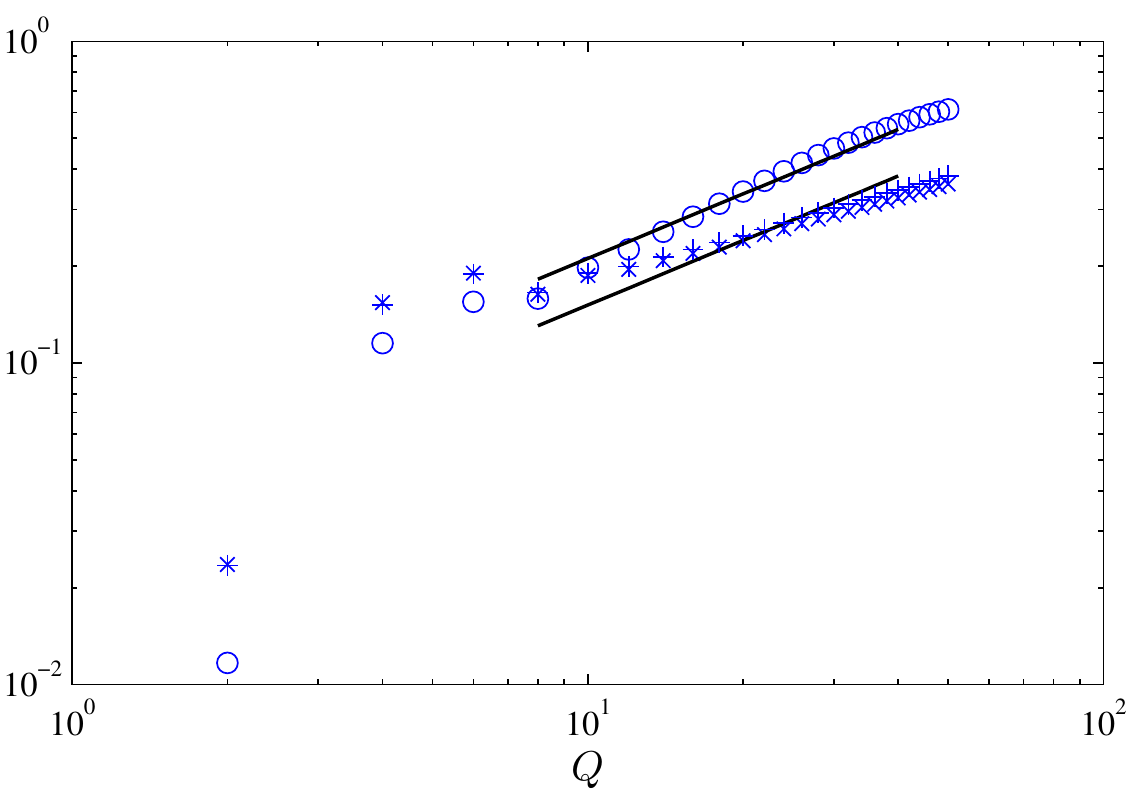}%,width=400pt,height=400pt } %, clip="[92 345 938 742]"}
\end{center}
\caption{ 
For $K=100$, we plot   $\langle|\Pi_K (\bu^{[Q]},\bu,\bu)|\rangle/\langle|\Pi_K|\rangle$ ($\circ$);   
    $\langle|\Pi_K (\bu,\bu,\bu^{[Q]})|\rangle/\langle|\Pi_K|\rangle$ ($\times$); 
and $\langle|\Pi_K (\bu,\bu^{[Q]},\bu)|\rangle/\langle|\Pi_K|\rangle$ ($+$)
using logarithmic bands $[Q/2,Q]$. 
The straight lines are for reference and have a $2/3$ slope. They extend over the fitting 
region which yields slopes of 0.77, 0.42, and 0.45 for ($\circ$), ($\times$), and ($+$),
respectively. Plots ($\times$) and ($+$) are almost identical, as expected.
}
\lb{FigureIR_point}
\end{figure}

\begin{figure}
\begin{center}
\epsfig{figure= 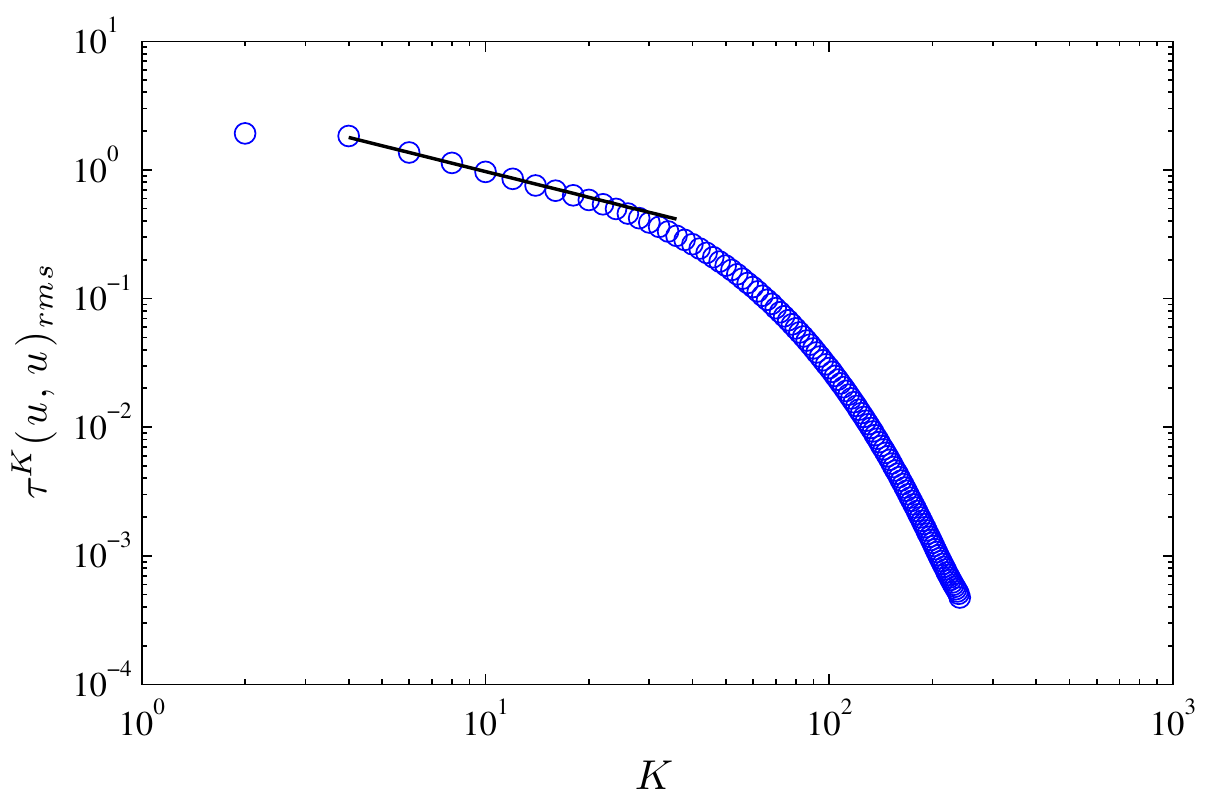}%,width=400pt,height=400pt } %, clip="[92 345 938 742]"}
\end{center}
\caption{ The $rms$ stress $\big(\btau_K(\bu,\bu)\big)_{rms} / E_{total}$ normalized by the energy 
of the system and plotted as a function of $K$. The straight line with -2/3 slope is for reference and 
extends over the fitting region which gives a slope of -0.71.
}
\lb{FigureStress}
\end{figure}

In Fig. \ref{FigureIR} we plot $-\langle\Pi_K(\bu^{[Q]},\bu,\bu) \rangle$ ($\circ$), 
$\langle\Pi_K(\bu,\bu,\bu^{[Q]}) \rangle$ ($\times$), and $\langle\Pi_K(\bu,\bu^{[Q]},\bu)\rangle$ 
($+$). Fitting with a power-law over the range $Q=8$ to $Q=40$, we obtain scaling 
laws of $Q^{1.08}$, $Q^{1.90},$ and $Q^{1.27},$ respectively,  
in fairly good agreement with our heuristic estimates $Q^{4/3}, Q^{7/3}$ and $Q^{4/3}.$
The IR decay rates are a little slower than those predicted, in agreement with the findings 
of Domaradzki \& Carati\cite{Domaradzki07b,Domaradzki09}. This can be attributed to 
the extreme shortness of the inertial range and the relatively smooth velocities at large scales 
in the simulation. Note that $\langle\Pi_K(\bu^{[Q]},\bu,\bu) \rangle <0$ over the range of the graph, 
becoming positive for $Q>60$. This can be understood in terms of detailed balance of energy in which 
the band of wavenumber modes $[Q]$ is receiving energy from modes $>K$, but the amount 
getting progressively smaller for $Q\ll K$ and more negligible compared to the positive 
total mean flux $\langle\Pi_K \rangle $.

\begin{figure}
\begin{center}
\epsfig{figure= 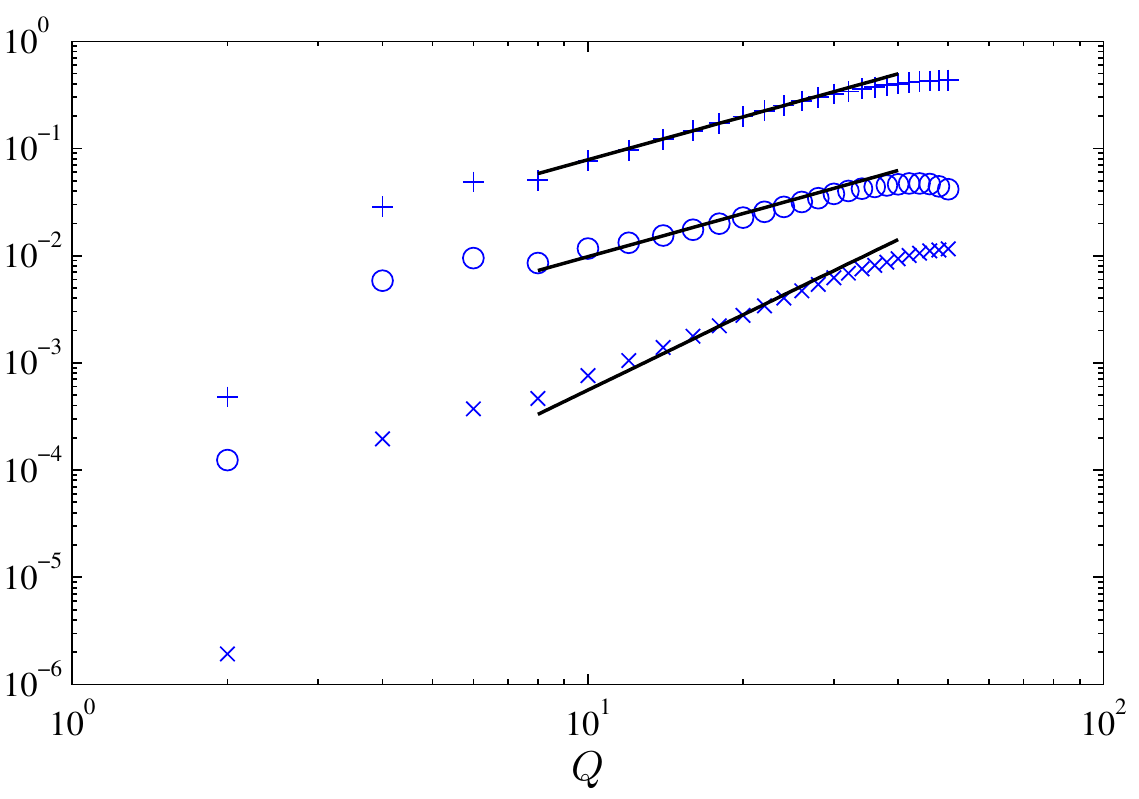}%,width=400pt,height=400pt } %, clip="[92 345 938 742]"}
\end{center}
\caption{ 
For $K=100$, we plot $-\langle\Pi_K (\bu^{[Q]},\bu,\bu) \rangle/\langle\Pi_K \rangle $ ($\circ$); 
 $\langle\Pi_K (\bu,\bu,\bu^{[Q]}) \rangle/\langle\Pi_K \rangle $ ($\times$); 
 and $\langle\Pi_K (\bu,\bu^{[Q]},\bu) \rangle/\langle\Pi_K \rangle$ ($+$)
using logarithmic bands $[Q/2,Q]$. 
The straight lines are for reference and have $4/3$ and $7/3$ slopes.
They extend over the fitting region which gives slopes of 1.08, 1.90, and 1.27 
for ($\circ$), ($\times$), and ($+$) respectively.
}
\lb{FigureIR}
\end{figure}

\section{Comparison with Previous Work}\lb{discuss}

Our entire analysis of the preceding section considered the SGS spectral energy flux (\ref{sgs}). 
It is more traditional to base discussions of wavenumber locality in turbulence on the triplet
transfer function 
$$\ol{T}(\bk,\bq,\bp)= {\rm Re}\,\{ik_i \langle
     \widehat{u}^*_j(\bk) \widehat{u}_i(\bq) \widehat{u}_j(\bp)\rangle\} 
     \delta_{\bk,\bp+\bq}. $$
This function can be interpreted as giving the mean rate of energy transfer into wavenumber 
mode $\bk$ from mode $\bp$ induced by mode $\bq$ and satisfies the ``detailed conservation'' 
property\cite{Onsager49}:
$$ \ol{T}(\bk,\bq,\bp)+ \ol{T}(\bk,-\bq,\bp)=0. $$
This transfer function is often integrated over spherical shells to define a quantity $\ol{T}(k,q,p)$
which depends only upon wavenumber magnitudes $k,p,q,$ restricted to values which 
can be assumed by the side-lengths of a closed triangle. In numerical simulations  
\cite{DomaradzkiRogallo, OhkitaniKida,Alexakisetal05b, Mininni06} this quantity is 
represented by 
\be T(K,Q,P)= \partial_i u^{[K-1,K]}_j u^{[Q-1,Q]}_i u^{[P-1,P]}_j,  
\lb{DNS-transfer} \ee
using band-pass velocity fields $\bu^{[K-1,K]}$ with bands of unit width. This function 
is defined pointwise in space, although usually its spatial average is taken over the 
flow domain. 

From the triplet transfer $\ol{T}(k,q,p)$ one can obtain the mean spectral flux as 
\be \ol{\Pi}_K=-\int_0^K dk \int_0^\infty dq \int_0^\infty dp\,\,\ol{T}(k,q,p), \lb{avrg-flux} \ee
or, pointwise in space, 
\be \Pi_K^{uns} =-\sum_{K'<K,Q',P'} T(K',Q',P')= -\partial_i u^{<K}_j (u_i u_j). \lb{ptws-flux} \ee
As discussed earlier, this quantity has the same space-average value as the 
SGS energy flux (\ref{sgs}). However, its pointwise properties are quite different,
since it is not Galilei-invariant unless averaged over space. Thus, the conventional
spectral flux (\ref{ptws-flux}), and also its variant form $-\partial_i u^{<K}_j (u_i u_j)^{<K}$, 
are pointwise scale-nonlocal objects, dominated by large-scale advection effects.
All such sweeping interactions, however, can be associated to space-transport of 
energy, with the SGS energy flux remaining as the unique pointwise Galilei-invariant 
measure of energy transfer to small-scales. See Eyink\cite{Eyink95,Eyink05} and I.
It was for this reason that we chose to analyze the spectral SGS flux in the 
previous section, since it enjoys much better scale-locality properties than the 
conventional spectral flux.  

On the other hand, the conventional spectral flux when averaged over space 
(or, assuming ergodicity, over time or over ensembles) is also dominated by local
triadic interactions. This was realized by the early pioneers of the 
subject\cite{Obukhov41a,Obukhov41b,Onsager45,Onsager49,Heisenberg48} and analyzed 
in detail by Kraichnan\cite{Kraichnan59,Kraichnan66,Kraichnan71} using 
closure approximations. A careful examination of Kraichnan's works shows 
that his arguments for scale-locality are based on the same basic ingredients 
as our rigorous proofs in the previous section. We shall briefly review here 
the asymptotic analysis in section 3 of Kraichnan (1966)\cite{Kraichnan66} 
using the ALHDIA closure, in order to discuss its relation to our arguments 
and also to facilitate later comparison with the numerical studies
\cite{BrasseurCorrsin,DomaradzkiRogallo,YeungBrasseur,OhkitaniKida,
Alexakisetal05b,Mininni06, Mininni08,Domaradzki07a,Domaradzki07b,
Domaradzki09}. In order to make direct contact between our proof and Kraichnan's 
analysis it is convenient to consider the quantity 
\be \ol{\Pi}_K(\bu,\bu^{[Q]},\bu)=
     -\int_0^K dk \int_{Q/2}^Q dq \int_{K}^\infty dp\,\,\ol{T}(k,q,p), \lb{flux-Kr1} \ee
which measures the mean energy flux  due to advection by modes in the 
wavenumber band $[Q].$ Kraichnan\cite{Kraichnan66} considered a slightly 
different set of quantities---in his notations, $T_{<Q}(K)$ and $\Pi_{<Q}(K)$---
or the transfer and flux due to advection by all modes with wavenumbers $<Q.$ 
He obtained asymptotic expressions for these in the limit $Q\ll K.$ However, 
his analysis and results carry over essentially to the flux quantity (\ref{flux-Kr1}). 
  
An important fact used by Kraichnan in his analysis is that, due to wavenumber 
conservation, this partial flux may be written as 
\be \ol{\Pi}_K(\bu,\bu^{[Q]},\bu)=
     - \int_{Q/2}^Q dq \int_{K-q}^K dk \int_{K}^{k+q} dp\,\,\ol{T}(k,q,p). \lb{flux-Kr2} \ee
Introducing dimensionless variables $u,v$ by $k=K-qv,\,p=K+qu,$ this may be 
rewritten as
\be \ol{\Pi}_K(\bu,\bu^{[Q]},\bu)=
     - \int_{Q/2}^Q dq \,q^2 \int_{0}^1 dv \int_{0}^{1-v} du\,\,\ol{T}(K-qv,q,K+qu). 
     \lb{flux-Kr3} \ee     
The result (\ref{flux-Kr3}) is exact, involving no approximation. It shows that the 
restriction on allowed wavenumbers interactions provides an additional small 
factor of $q^2$ for small $q,$ which proves essential to obtain IR spectral locality.  

To proceed further requires an explicit expression for the transfer function.    
Kraichnan's ALHDIA closure yields the formula
\be \ol{T}(k,q,p,t) \equiv \hat{B}_{kpq} \Delta_{kpq} \int_{t_0}^{t} ds 
       \big[ G(k; t|s) U(p; t|s) -  G(p; t|s) U(k; t|s) \big] U(q; t|s); 
\lb{Kr-transfer}\ee
see equation (2.6) of Kraichnan (1966)\cite{Kraichnan66}. Here $U(k;t|s)$ and 
$G(k;t|s)$ are 2-time Lagrangian velocity-correlation and mean-response functions.
The factor $\Delta_{kpq}$ is equal to 1 if the wavenumbers $k,p,q$ can be
the sides of a closed triangle and is equal to 0 otherwise.  Finally, the factor 
$\hat{B}_{kpq}=4\pi^2 k^2p^2q(xy+z^3),$ where $x=\cos\alpha,y=\cos\beta,
z=\cos\gamma$ are the cosines of the angles $\alpha,\beta,\gamma$ that are 
opposite to sides of length $k,p,q$ in the wavenumber triangle. Substituting 
the steady-state, inertial-range scaling forms 
\be G(k;t|s)=g(\varepsilon^{1/3}k^{2/3}(t-s)),\,\,\,\,
U(k;t|s)=\frac{C}{2\pi}\varepsilon^{2/3}k^{-11/3}r(\varepsilon^{1/3}k^{2/3}(t-s)), 
\lb{Kr-GU} \ee
one obtains the asymptotic expression for $k=K-qv,\,p=K+qu,$ and $q\ll K$
\be \ol{T}(k,q,p) \simeq  -\varepsilon w(1-w^2) C^2 I K^{-4/3} q^{-5/3}, 
\lb{asympt-Kr} \ee
where $w=u+v$ and $I$ is an integral over the scaling functions $g(\tau),r(\tau).$ 
See Kraichnan\cite{Kraichnan66} and Appendix B for details. We just note 
here one signficant feature of the calculation, which is the near cancellation 
between the two input and output terms in (\ref{Kr-transfer}) for $k\approx p
\approx K.$ The physics of this was discussed by Kraichnan\cite{Kraichnan66}, 
p.1733:
\begin{quote}
``These terms separately give contributions proportional to the energy in 
the wavenumbers $<q,$ not to the mean-square vorticity. ...The input and
output contributions from low wavenumbers are proportional to energy because
they represent convection as well as straining. Convection of high-wavenumber
structures by strongly excited low-wavenumber velocity components implies
a rapid exchange of phase of the high-wavenumber Fourier amplitudes....
This exchange is represented in (2.6) [our (\ref{Kr-transfer})] by the large,
cancelling input and output contributions. The net contribution $T_{<q}(k,t)$ 
represents the effect of straining alone.'' 
\end{quote}
These remarks are consistent with our observation that the spectral energy flux 
$\Pi_K^{uns}$ in (\ref{ptws-flux}) is pointwise non-local and dominated by convective 
sweeping, with such effects only cancelling in the average over space. 
The cancellation between input and output terms for $k\approx p\approx K$ 
also yields an additional small factor of $q$ crucial to the final result.  

Substituting the closure approximation (\ref{asympt-Kr}) into the expression 
(\ref{flux-Kr3}) yields finally that 
\be \ol{\Pi}_K(\bu,\bu^{[Q]},\bu)\simeq {\mathcal C}\varepsilon \left(\frac{Q}{K}\right)^{4/3},
       \,\,\,\,\,\,\,\,\, Q\ll K, \lb{Kr-4third} \ee
for ${\mathcal C}=\frac{1}{10}C^2I\big(1-\frac{1}{2^{4/3}}\big)\doteq 0.503.$ For details,
see Appendix B. We thus recover Kraichnan's 4/3 scaling law for decay of IR-nonlocal 
contributions. The factor $Q^{4/3}$ arises from integration of  the product $q^{1/3}=
q^2q^{-5/3}$ over the octave band $[Q/2,Q].$ Just as in our rigorous proof in the 
preceding section, there is a competition between the decaying factor $q^2$ which 
arises from the restriction on wavenumber interactions and the growing factor 
$q^{-5/3}$ which arises from the increase in energy of low-wavenumber modes. 
For a similar calculation and analysis, see Verma et al.\cite{Vermaetal}, section 3. 

We have focused so far on energy flux but many studies \cite{DomaradzkiRogallo,
OhkitaniKida,Alexakisetal05b,Mininni06, Mininni08,Domaradzki07a,Domaradzki07b,
Domaradzki09} instead consider energy transfer between wavenumber bands.
Our results also imply scale-locality of  suitable transfers.  Note, for example, that 
the partial flux in (\ref{flux-Kr2}) for $Q<K/2$ equals
\begin{eqnarray}
\ol{\Pi}_K(\bu,\bu^{[Q]},\bu) & = & 
        \ol{\Pi}_K(\bu^{[K]},\bu^{[Q]},\bu^{[P]})  \cr
     & = & - \int_{K/2}^K dk \int_{Q/2}^Q dq \int_{P/2}^{P} dp\,\,\ol{T}(k,q,p), \cr
     & \equiv & -\ol{T}([K],[Q],[P]), 
\lb{flux-Kr4} \end{eqnarray} 
with $P=2K.$ In the last line we have defined the mean triplet transfer function with 
octave bands $[K]=[K/2,K].$ Because this quantity equals the partial flux, our previous 
bounds and scaling results all apply, showing that this quantity for $Q\ll K$ is negligible 
compared with the total transfer rate from adjacent band interactions.  

It is very important for the closure calculation presented in this section and also for the 
rigorous proof in the preceding section that such logarithmic bands be employed. When
demonstrating IR locality in the advecting mode, we have used the fact that modes with 
wavenumbers $Q\ll K$ interact with a very restricted subset of the modes contained in 
$\bu^{[P]}$, resulting in an overall weak contribution to the flux. On the other hand, if linear 
bands are used, the modes contributing to $\bu^{[P-1,P]}$ are already restricted and, thus, 
taking $Q$ smaller does not impose any further restriction. Thus, the space-average of 
$T(K,Q,P)$ in (\ref{DNS-transfer}) must increase for smaller $Q$ simply because 
$\bu^{[Q]}$ gets bigger at  the larger scales, where most of the energy resides. This 
can be verified directly from the asymptotic formula (\ref{asympt-Kr}) of the ALHDIA  
closure for $\ol{T}(k,q,p),$ which is essentially the same quantity.  However, $T(K,Q,P)$ 
and other such objects which compare single triads vastly underestimate the 
contribution from the local triads, which, when taken into account cumulatively, 
swamp the non-local interactions through sheer number.

Such remarks apply to studies which have claimed to verify a cascade mechanism of 
``local transfer by nonlocal triads'' in numerical simulations. For example, several 
such works\cite{DomaradzkiRogallo, OhkitaniKida, Alexakisetal05b, Mininni06} have 
calculated the partially-summed transfer  
$$ \ol{T}(K,P) = \sum_Q \ol{T}(K,Q,P) =
\frac{1}{V} \int d^3x\,\,\partial_i u_j^{[K-1,K]} u_i u_j^{[P-1,P]}. $$
This can be interpreted as the mean rate of energy transfer into wavenumber band $[K-1,K]$ 
from band $[P-1,P],$ mediated by all other wavenumbers.  The cited numerical studies
have found that the energy transfer to the $K$th band has largest positive contribution from 
$P=K-k_f$ and largest negative contribution from $P=K+k_f$, mediated by modes at 
the forcing wavenumber $k_f$. This result was suggested to support the dominance of 
nonlocal triads in energy cascade\cite{DomaradzkiRogallo, OhkitaniKida, Alexakisetal05b, 
Mininni06}.  In fact, the result should have been expected and is  completely consistent 
with scale-local interactions! Notice that this behavior is predicted, for example, by the 
asymptotic formula (\ref{asympt-Kr}) from Kraichnan's ALHDIA closure, since its integral 
over $q$ is clearly dominated by the peak of the $q^{-5/3}$ spectrum at $q=k_f.$ However, 
Kraichnan correctly deduced from this expression that cascade proceeds by the local triads, 
not the non-local triads.  This can be seen by considering instead the similar quantity
$$ \ol{T}([K],[P]) = \frac{1}{V} \int d^3x\,\,\partial_i u_j^{[K]} u_i u_j^{[P]}, $$
employing octave bands. Our previous discussion applies to this object, showing that 
the non-local triads make a negligible contribution. 

To underline this conclusion, we shall exhibit illustrative numerical results from
a $512^3$ simulation using hyperviscosity \cite{BorueOrszag95}:
\be  \partial_{t} \bu +  (\bu\cdot\grad)\bu = -\grad p + \nu_{10}(\grad^2)^{5} \bu + {\bf f} \ee
We employed the same forcing (\ref{forcing}) as for our viscous simulations but with 
$k_f = 1,$ and we took hyperviscous coefficient $\nu_{10} = 2\times10^{-21}$. 
%The scaling results were consistent with those from the data generated using regular viscosity, but
%there was no net gain in the range of inertial scaling of the spectrum due to a wide bottleneck \cite{Falkovich} known
% to become more pronounced with the use of hyperviscosity \cite{Biskampetal98,HaugenBrandenburg}. 
These choices were made to achieve a constant mean energy flux $\ol{\Pi}_K=\varepsilon$ 
over as long a range of wavenumbers as possible. See Figure \ref{FigureCascade}a.  Just
below, in Figure \ref{FigureCascade}b, we plot $\ol{T}(K,P)$  for linear bins magnified by 
a factor of 10 (dashed line) and $\ol{T}([K],[P])$ for octave bands (solid line). Both quantities 
are plotted as functions of $P$ for $K=50$ and have been normalized by the mean flux 
$\varepsilon.$ 

The results for unit-width bands agree with those of the earlier 
studies\cite{DomaradzkiRogallo, OhkitaniKida, Alexakisetal05b, Mininni06}, 
while those for octave bands are similar to those presented by Verma et al.
\cite{Vermaetal}, Figure 5 for logarithmic bands (base-$\sqrt{2}$ rather than our base-$2$). 
However, it is quite instructive to plot the two quantities together. The result for $\ol{T}(K,P)$
illustrates the contribution of  ``local transfer by nonlocal triads'', with the familiar 
feature of peaks separated by $2k_f.$ Notice, however, that the magnitude is 
miniscule compared both with the mean flux $\varepsilon$ and also with the 
summed effect of all the local triads, which dominates in $\ol{T}([K],[P]).$
This kind of transfer will become an even smaller fraction of the total for larger 
$K,$ decreasing as $\sim K^{-4/3}$ according to the Kraichnan estimate 
(\ref{Kr-4third}). The plot of $\ol{T}([K],[P])$ shows that transfer into $[K]$ 
has largest positive contribution from the octave band $[P]$ with $P\approx K/2$ 
and largest negative contribution from the band with $P\approx 2K.$ This 
demonstrates the multiplicative, self-similar nature of the local cascade. 

%According to our estimates, the amount of energy entering 
%$[K]$ from bands $P\ll K$ should decay like $P^{4/3}$ and  the solid-line plot in figure 
%\ref{FigureCascade}b has a $P^{1.12}$ decay over the range $P\in[1,30]$. We also 
%expect that the amount of energy cascading to higher wavenumbers would decay like 
%$P^{-4/3}$ but such a scaling, which we have shown in figure \ref{FigureUV}, does not 
%appear in figure \ref{FigureCascade}b because $K=50$ is not far enough from the 
%dissipation range.

%\newpage

%This is why $\int d\bx \Pi_K(\bu,\bu^{[P]},\bu)$ was found to get smaller for smaller $P$, 
%which restricts the number of modes contributing to the strain due to wavenumber conservation. 
%In light of this, we can understand why several studies which have used linear bands like $[K-1,K]$ 
% have concluded that the energy transfer in turbulence is non-local. In fact, the quantity
%$$ \frac{ \partial_j u^{[K-P,K]}_i u^{>K}_i u^{[P]}_j}{ \partial_j u^{[K]}_i u^{>K}_i u^{>K}_j} 
%\sim \frac{P^{-h}K^{-2h}}{K^{-3h}}$$
%if linear bands are used, which attains its maximum when $\frac{P}{K} \to 0$. On the other hand, 
%if logarithmic bands are used, then we have argued that the above ratio
%becomes vanishingly small like $\sim \big(\frac{P}{K}\big)^{1-h}$ without even considering 
%decorrelation effects from space integration.
\begin{figure}
\begin{center}
\epsfig{figure= 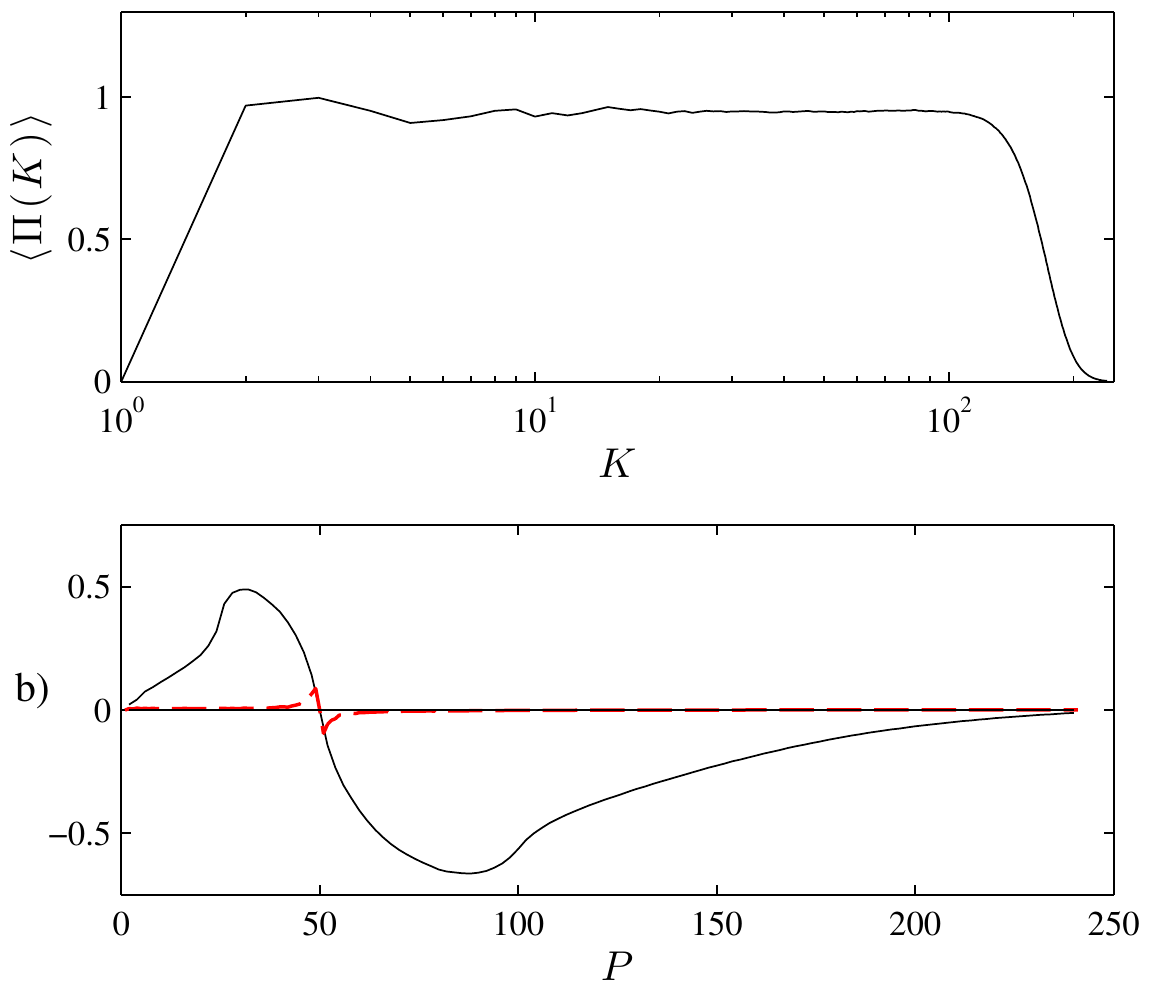}%,width=400pt,height=400pt } %, clip="[92 345 938 742]"}
\end{center}
\caption{ Using data from the hyperviscous simulation, we plot (a) the energy flux $\langle \Pi_K \rangle$.
In (b), we plot, for $K=50$, as a function of $P$:
(solid) $\langle\partial_i u_j^{[K]}u_iu_j^{[P]}\rangle/\langle\Pi_K \rangle $ using octave bands $[K/2,K],\,
[P/2,P]$, and (dashed) $10\times \langle\partial_i u_j^{[K-1,K]}u_iu_j^{[P-1,P]}\rangle/\langle\Pi_K \rangle $ 
using linear bands. We multiply the latter by $10$ for comparison. The dashed-line plot 
shows energy transfer in incremental steps with the distance between peaks being $2$, which is twice 
the forcing wavenumber, $k_f=1$.This kind of transfer represents a negligible fraction of the flux 
$\langle\Pi_K \rangle $, as the plot shows, which will decrease even more ($\sim K^{-4/3}$) for 
larger $K$. The solid-line plot has peaks at $P=32$ and $88$, showing transfer in multiplicative steps. 
The solid-line decays like $P^{1.12}$ over the range $[1,30]$, quite close to the expected power $P^{4/3}$.
There is no similar discernible power-law decay in the ultraviolet limit of large $P$ due to dissipation.
}
\lb{FigureCascade}
\end{figure}

Our results rigorously disprove a number of suggestions in the recent literature. Alexakis et al.
(2005)\cite{Alexakisetal05b} and Mininni et al. (2006,2008)\cite{Mininni06, Mininni08} have claimed
on the basis of $1024^3$ and $2048^3$ DNS that the ratio $\ol{\Pi}_K(\bu,\bu^{[Q]},\bu)/\ol{\Pi}_K$ with 
$Q\simeq k_F$ asymptotes to a value of about $0.10$-$0.20$ for large $K$ in the inertial range.
A long plateau within the inertial range is ruled out by our rigorous estimate (\ref{main}) which shows that 
this quantity is bounded at least as $O(\varepsilon (KL)^{-\alpha})$ for large $K,$ with $\alpha=\sigma_2
+\sigma_4-1/2\doteq 1/6.$ According to the estimate (\ref{Kr-4third}) from Kraichnan's 
ALHDIA closure theory, the decay rate is actually $\sim \varepsilon (KL)^{-4/3},$ which is even 
faster.  Whatever is the explanation of the plateau observed in the reported simulations
\cite{Alexakisetal05b,Mininni06,Mininni08} it cannot be an inertial range effect.  Indeed, 
the asymptote observed in their numerical data occurs at high wavenumbers $K$ beyond
the constant flux inertial range and within the pronounced ``bottleneck'' of their energy spectrum.
%Figure 3 of Mininni et al. (2006) \cite{Mininni06} and Figure 1 of Mininni et al. (2008) \cite {Mininni08}. 
A Reynolds-number dependence of the plateau level has furthermore been reported 
in the most recent paper\cite{Mininni08}(see Fig.4 there)
$$ \frac{\ol{\Pi}_{K_\lambda}(\bu,\bu^{[k_F]},\bu)}{\ol{\Pi}_{K_\lambda}}
     \sim  Re^{-\beta},\,\,\,\,\, \mbox{$\beta\doteq$ 0.6\textendash 0.7}, $$
where $K_\lambda$ is the Taylor-scale wavenumber. Since $L/\lambda\simeq Re^{1/2},$
this is completely consistent with Kraichnan's estimate (\ref{Kr-4third}), which would yield a  
scaling law $Re^{-2/3}$ for the above ratio. Inertial-range theory cannot explain, however, 
why the fraction stays constant for $K>K_\lambda.$  We conjecture that this is an effect of 
the bottleneck.  In Figure \ref{FigPouquet} we plot the same ratio using data from our 
hyperviscous simulation. It shows a decay very close to $K^{-4/3}$ through the inertial range, 
until it asymptotes at  $2$-$3\%$ over the region of the bottleneck, which is known to be
more prominent for hyperviscosity \cite{Biskampetal98,HaugenBrandenburg}. Note that 
we do not observe a similar plateau in our simulation with regular viscosity, which does not 
have a conspicuous bottleneck (see Fig.\ref{FigureSpectrum}). 

\begin{figure}
\begin{center}
\epsfig{figure= 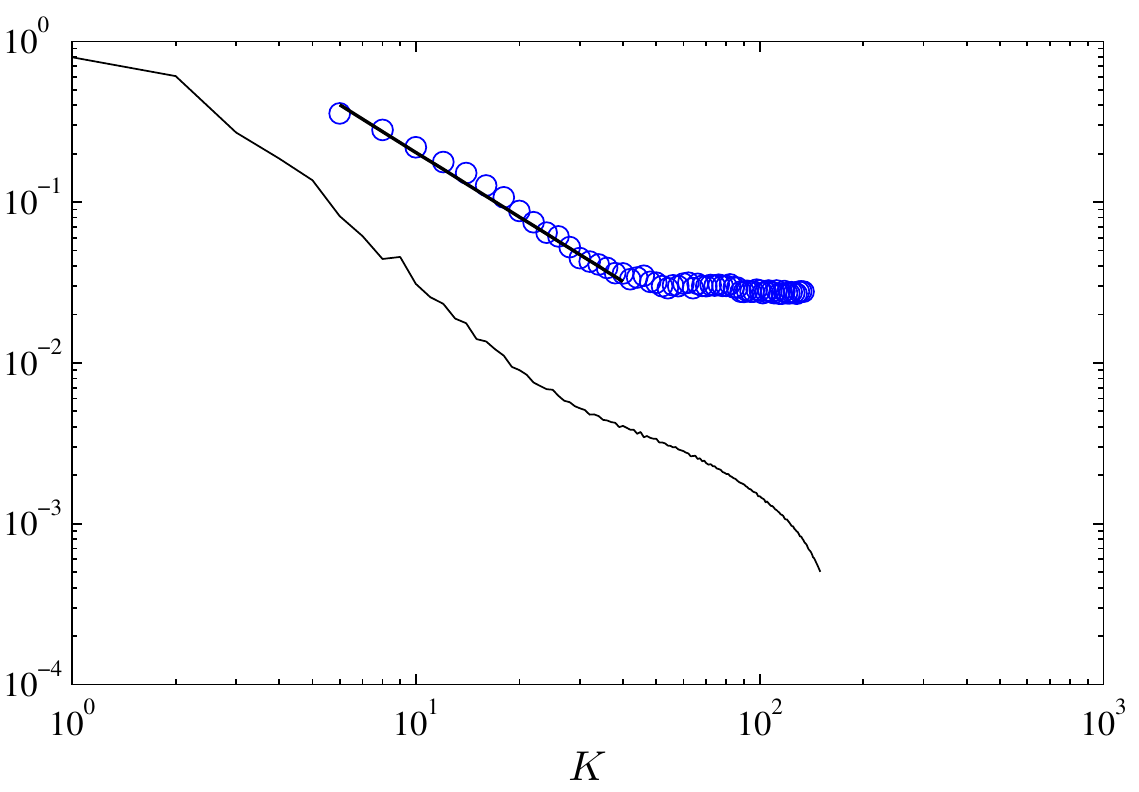}%,width=400pt,height=400pt } %, clip="[92 345 938 742]"}
\end{center}
\caption{ Using data from the hyperviscous simulation, we plot 
$\langle \Pi_K(\bu,\bu^{<6},\bu) \rangle/ \langle \Pi_K\rangle$ ($\circ$) as a function of $K$.
The straight line with slope  $-4/3$ is for reference and extends over the fitting range
of the graph, which gives a decay rate of $K^{-1.32}$. Beyond this range, the plot
asymptotes at $2$-$3\%$. The plot of the energy spectrum (solid line) suggests  that this 
trend is related to the bottleneck in the crossover between the inertial and dissipation 
ranges of the spectrum.}
\lb{FigPouquet}
\end{figure}

Alexakis et al. (2005)\cite{Alexakisetal05b} 
and Mininni et al. (2006,2008)
\cite{Mininni06,Mininni08} have further speculated that their numerical results---which they 
interpreted to show the dominance of non-local interactions to mean transfer---may be due to 
intermittency and long-lived vortical structures. However, the bounds on {\it space-average} 
energy flux in the current work are little affected by intermittency, since they involve only 
low-order scaling exponents $\sigma_p$ with $p= 2\sim 4.$ Our estimates for large $p$ and 
the pointwise estimates for graded filters\cite{Eyink05,EyinkAluie} will indeed be modified by intermittency, 
which has the effect of {\it improving} the IR estimates and degrading the UV estimates. UV 
locality will nevertheless hold at every space point with a positive H\"{o}lder exponent 
of the velocity field.

\section{Conclusions}

We have proved by rigorous estimates that spectral energy flux in three-dimensional turbulence
is dominated  by local wavenumber triads. No additional averaging over scales,  as in 
Eyink (1994)\cite{Eyink94}, was required. Furthermore, we showed that spectral locality
holds without additional cancellations resulting from space averaging, which, 
when included, further decrease the net non-local contribution. Our proof exploited 
four main ingredients: (1) The SGS spectral energy flux, which is the unique Galilei-invariant 
measure of the energy flux across scales. It has the sweeping effects subtracted, which is essential 
to prove infrared locality. (2) Scaling properties that are observed empirically to hold for the turbulent 
velocity field. In particular, we employ the strong scaling condition eq.(\ref{spectrum}) on the energy 
spectrum, which rules out the mathematical counterexample of Eyink\cite{Eyink94}. (3) Wavenumber
conservation,  which restricts the range of interactions in Fourier space. (4) Logarithmic 
bands of wavenumber modes which, unlike linear bands, represent the cumulative effect 
of the growing number of local triads.

We were thus able to explain the results of previous research claiming the dominance 
of non-local interactions in the energy cascade \cite{DomaradzkiRogallo,OhkitaniKida,
Alexakisetal05b,Mininni06}. Since these studies used linear bands in decomposing the velocity 
field,  they were effectively comparing triads of \emph{single} Fourier modes with each other. 
The most dominant of these triads are the ones with a large-scale advecting mode. However, 
such non-local triads are few, an idea implicit in the work of Kraichnan (1966)\cite{Kraichnan66} 
and emphasized more recently by Verma et al. (2005) \cite{Vermaetal}. Thus, their effect becomes 
more negligible as the energy cascades to smaller scales, whereas the number of local triads grows
geometrically, swamping the non-local effects. Therefore, the Kolmogorov picture of local cascade
is consistent with the presence of non-local triadic interactions, since these become vanishingly 
insignificant in a longer inertial-range. 

We find that the scale-locality properties of the SGS flux with a sharp spectral filter are close to those 
with graded filters established by Eyink\cite{Eyink95,Eyink05} and in paper I. Our numerical results for the 
{\it space-average} of the SGS spectral flux are essentially the same as those for graded filters, 
in agreement with the earlier findings of Domaradzki \& Carati\cite{Domaradzki07a,Domaradzki07b}.
The locality properties of the SGS spectral flux $\Pi_K$ are possibly not quite as good {\it pointwise}
as those of SGS flux  with graded filters. We could prove some nearly uniform estimates on UV 
and IR locality, but with decay rates that are probably sub-optimal and with additional $K$-dependent 
factors. More seriously, we could not prove pointwise estimates on the SGS spectral flux involving 
the local H\"{o}lder exponent at the space point $\bx$. 
% Wavelets: tools for science & technology
% By StŽphane Jaffard, Yves Meyer, Robert Dean Ryan
% SIAM, 2001
However, we would like to moderate the claim of Eyink\cite{Eyink05} that spectral energy
flux ``is an inappropriate measure of energy transfer.'' This is true for the {\it unsubtracted} flux 
$\Pi_K^{uns},$ which is pointwise non-Galilei-invariant and becomes scale-local only when 
averaged over space. On the other hand, the SGS spectral flux $\Pi_K$ is scale-local, even
in the absolute sense, without cancellations due to space-averaging. In particular, the 
sharp spectral filter has a sound theoretical basis for use in large-eddy simulation (LES) 
modeling of turbulent energy cascade. 

The width of wavenumber bands is more important for scale-locality than the grading 
of the filter kernel. As we have seen, transfer functions for individual wavenumber triads 
or transfer functions with wavenumber bands of fixed width on a linear scale may have 
the greatest magnitude for non-local triads. The situation is completely different for transfer 
functions with bands of fixed width on a logarithmic scale. The transfers for such \emph{logarithmic 
bands} are dominated by local triad interactions, which, by their vastly greater numbers, 
overwhelm the contribution of the non-local triads. Only the collective contribution 
of the local triads is large enough to explain the observed flux of energy to  small 
scales.  In addition, it is only quantities summed over logarithmic bands which can be  
simultaneously localized both in scale and in space. This is a basic fact of harmonic analysis, equivalent 
to the Heisenberg uncertainty principle\cite{Jaffardetal01}. To constitute an ``eddy'' of half the size 
requires twice as many wavenumber modes to achieve the necessary localization in space. 
If one believes, as we do, that the essential dynamics of turbulent cascade involves modes resolved 
both in scale and in space, then summation over logarithmic wavenumber bands is necessary to its 
physical description. 
%The picture of energy cascade as ``deformation work'' of large-scale strain acting 
%locally in space against small-scale stress \cite{TL} depends upon such a device.

Fourier series, without a doubt, are often a useful tool to analyze turbulent cascades. However, individual 
Fourier modes have no physical significance and it is dangerous  to think of the elementary interactions 
as those involving individual wavenumber triads. A single Fourier mode is a wave which extends periodically  
throughout all space, whereas incompressible fluids (when constant-density, non-rotating, etc.) exhibit  
no linear wave phenomena. Consider the case of turbulent flow in a wall-bounded domain, like air 
in a room or water in a cup. The velocity field for such a flow may be decomposed into Fourier modes 
by imbedding the flow domain in a periodic box. A single triad then represents interacting periodic 
waves, exchanging energy beyond the solid walls of the domain! This clearly has no physical reality.
Fourier modes are only one convenient basis, among many, that may be used in the scale decomposition 
of the velocity field, and the physics cannot depend on the expansion coefficients of any particular 
basis. Only the sum over the basis elements can have intrinsic meaning. 

We have proved in this work that, asymptotically for long inertial ranges, the energy cascade proceeds neither
in large strides between widely separated Fourier bands nor in small incremental wavenumber steps.
Rather, it proceeds in multiplicative steps between Fourier bands which are adjacent to each other 
on a logarithmic scale.  However, the bounds on the non-local triad contributions to energy flux 
decay only as power-laws, like the 4/3-power law of Kraichnan.  The cascade process can thus 
be accurately described as ``diffuse'' \cite{Kraichnan59} or ``leaky'' \cite{TL}. Very large Reynold's 
numbers are needed,  larger than what today's numerical simulations are able to attain,  in order 
for locality and its consequences to become fully manifested.

\vspace{.5in} 
\noindent {\small
{\bf Acknowledgements.} H.A. wishes to thank Shiyi Chen, Minping Wan and Dmitry Shapovalov 
for assistance in developing the simulation code used in this work.  Computer time was provided 
by the Digital Laboratory for Multi-Scale Science at the Johns Hopkins University. This work was 
supported by NSF grant \# ASE-0428325 at the Johns Hopkins University.}

%=============================================================
%\begin{center}
%\includegraphics[totalheight=0.72\textheight,viewport=70 190 552 750,clip]{Fisher_Abstract-FST08.pdf}
%\end{center}
%scale=0.65]
%=============================================================

\newpage
\renewcommand{\theequation}{A-\arabic{equation}}
\setcounter{equation}{0}  % reset counter 
\setcounter{Prop}{0}  % reset counter 
\section*{APPENDIX A}  % use *-form to suppress numbering

We give here more details and precise statements of our rigorous results. We first  explain 
briefly the justification of the scaling relations in eq. (\ref{scaling}) that are employed in our
argument. A basic observation of experiments and numerical simulations \cite{Anselmetetal,Chenetal,Sreenivasanetal} on turbulent flow is the scaling of structure 
functions of velocity-increments $\delta\bu(\br;\bx)=\bu(\bx+\br)-\bu(\bx):$ 
\be \big\langle|\delta \bu(\br)|^p \big\rangle^{1/p}  \sim  A_p r^{\sigma_p} \lb{structure} \ee
for $\eta_p\ll r\ll L$ with $\eta_p$ a short-distance viscous length-scale. As earlier, we shall 
interpret $\langle\cdot\rangle$ as a space-average. We assume that the scaling range becomes 
longer as $\nu\rightarrow 0,$ so that $\eta_p\rightarrow 0.$ A weaker form of ``scaling'' implied 
by (\ref{structure}) is then that 
\be \sigma_p =  \liminf_{r\rightarrow 0}\frac{\log \big\| \delta \bu(\br) \big\|_p}{\log r} \lb{besov} \ee 
with $\|\cdot\|_p=\langle |.|^p\rangle^{1/p}$ the $L_p$-norm. 
If we assume also that $\langle|\bu|^p\rangle<\infty$, then relation (\ref{besov}) with $0<\sigma_p<1$
implies that $\bu$ is {\it Besov regular} with maximal Besov exponent $\sigma_p$ of order $p.$
That is, $\bu\in B^{s,\infty}_p(\mathbb{T}^3)$ for all $s<\sigma_p$ but for no $s>\sigma_p.$ 
E.g. see \cite{Triebel,Eyink95,BasdevantPerrier96} for more background on Besov spaces
and their relations to turbulent structure functions. 
%V. Perrier and C. Basdevant, ``Besov norms in terms of the continuous wavelet transform . Application 
%to structure functions,'' Math. Models \& Meth. Appl. Sci. {\bf 6} 649--664 (1996).

Assuming validity of (\ref{structure}) [or even (\ref{besov})], a Paley-Littlewood criterion for 
Besov regularity implies a rigorous version of the second scaling relation in eq.(\ref{scaling}): 
\be \sigma_p =  \liminf_{K\rightarrow\infty}\frac{\log \|\bu^{[K]}\|_p}{\log(1/K)}, \lb{besov-PL} \ee
with $\bu^{[K]}(\bx)=\sum_{K/2<|\bk|<K}e^{i\bk\bdot\bx} \widehat{\bu}(\bk).$ See 
Triebel\cite{Triebel}, Theorem 3.5.3(i). Here the norm on wavenumbers must 
be taken to be $|\bk|_\infty=\max\{k_x,k_y,k_z \}$ or $|\bk|_1=|k_x|+|k_y|+|k_z|$ and 
not the Euclidean norm $|\bk|_2=(k_x^2+k_y^2+k_z^2)^{1/2}.$ As we shall see below,
other delicate aspects of Fourier analysis also rule out the use of the Euclidean norm 
for a rigorous proof of locality bounds. Note that (\ref{besov-PL}) is actually more general than 
(\ref{besov}) and gives the maximal Besov exponent of $\bu$ for any $\sigma_p\in\mathbb{R}$ 
and $p\geq 1.$ The first scaling relation in eq.(\ref{scaling}) has a similar rigorous statement 
for $\sigma_p>0,$
\be \sigma_p =  \liminf_{K\rightarrow\infty}\frac{\log \|\bu^{>K}\|_p}{\log(1/K)}, \lb{besov-approx} \ee
which follows from results on the strong summability of Fourier partial sums
that characterize Besov spaces. See Triebel\cite{Triebel}, section 3.73, Theorem 1(i). 
Finally, $\bu\in B^{s,\infty}_p
(\mathbb{T}^3)$ iff $\grad\bu\in B^{s-1,\infty}_p(\mathbb{T}^3)$ and then the result
\be \sigma_p-1 =  \liminf_{K\rightarrow\infty}\frac{\log \|\grad\bu^{<K}\|_p}{\log(1/K)}
      \lb{besov-grad} \ee
for $\sigma_p<1$ follows from (\ref{besov-PL}). 

We now state and prove a rigorous analogue of inequality (\ref{UV-local}), implying
UV locality of the spectral SGS flux:
\begin{Prop} If $\bu$  has maximal Besov index $\sigma_{3p}$ of order $3p$ for $p\geq 1,$ 
then 
$$ \|\Pi_K (\bu,\bu,\bu^{[Q]})\|_p \le C_p K^{1-\sigma} Q^{-2\sigma}$$
when $Q>4K,$ for any $\sigma<\sigma_{3p}.$
\end{Prop}

\noindent {\it Proof of Proposition 1:} 
The starting point of the argument is the identity
$$  \Pi_K(\bu,\bu,\bu^{[Q]}) =  -\partial_i u_j^{<K}    \big(u_i^{[\frac{Q}{2}-K,Q+K]} u_j^{[Q]}\big)^{<K}. $$
derived in the text. Then by the H\"{o}lder inequality 
$$ \|\Pi_K (\bu,\bu,\bu^{[Q]})\|_p \le 
\|\grad\bu^{<K}\|_{3p} \|\big( \bu^{[\frac{Q}{2}-K,Q+K]}\bu^{[Q]}\big)^{<K}\|_{3p/2} $$
We next exploit the $L_p$-boundedness  of Fourier multipliers for polygonal partial 
summation to obtain
$$ \|\Pi_K (\bu,\bu,\bu^{[Q]})\|_p \le 
 C\|\grad\bu^{<K}\|_{3p} \|\bu^{[\frac{Q}{2}-K,Q+K]}\bu^{[Q]}\|_{3p/2} $$
for some constant $C>0.$ See e.g., Krantz\cite{Krantz},Theorem 3.4.5. Note that such 
an estimate cannot be obtained with the Euclidean norm on wavenumbers, due to the 
failure of boundedness of Fourier multipliers for the ball. See Fefferman (1971) \cite{Fefferman}. 
However, we can approximate the sphere with any convex polygon with an arbitrary but fixed 
number of sides and our rigorous results will hold (see, for example, Krantz\cite{Krantz}, and 
other references on the subject \cite{Sogge,SteinWeiss}).  Using again the H\"{o}lder inequality 
gives
$$ \|\Pi_K (\bu,\bu,\bu^{[Q]})\|_p \le 
 C\|\grad\bu^{<K}\|_{3p} \|\bu^{[\frac{Q}{2}-K,Q+K]}\|_{3p}\|\bu^{[Q]}\|_{3p}. $$
Writing $\bu^{[\frac{Q}{2}-K,Q+K]}=\bu^{>\frac{Q}{2}-K}-\bu^{>Q+K},$ we can use (\ref{besov-approx})
to obtain the bound 
$$ \|\bu^{[\frac{Q}{2}-K,Q+K]}\|_{3p}\leq \|\bu^{>\frac{Q}{2}-K}\|_{3p}+\|\bu^{>Q+K}\|_{3p}
      \leq C_p' \left[ (Q-2K)^{-\sigma}+(Q+K)^{-\sigma}\right] \leq C_p'' Q^{-\sigma} $$
for any $\sigma<\sigma_{3p}$ and $Q>4K.$ Using the similar estimates for $\sigma<\sigma_{3p}$
$$ \|\grad\bu^{<K}\|_{3p} <C_p' K^{1-\sigma},\,\,\,\,\,\|\bu^{[Q]}\|_{3p}\leq C_p' Q^{-\sigma} $$
from (\ref{besov-grad}) and (\ref{besov-PL}) yields the final inequality. \hfill $\Box$

Finally, we state and prove a rigorous analogue of inequality (\ref{main}), implying
IR locality of the spectral SGS flux: 
\begin{Prop} If $\bu$ has maximal Besov indices $\sigma_2$ and $\sigma_4$ of order 
${\rm 2}$ and ${\rm 4}$, resp., and if the strong scaling relation (\ref{spectrum}) holds 
for the energy spectrum, then
$$\|\Pi_K (\bu,\bu^{[P]},\bu)\|_1 \le C K^{1/2-\sigma'-\sigma} P^{1/2-\sigma'}$$
when $P<K/2,$ for any $\sigma\leq \sigma_2$ and any $\sigma'<\sigma_4.$
\end{Prop}

\noindent {\it Proof of Proposition 2:}
We begin with the identity
$$  \Pi_K(\bu,\bu^{[P]},\bu) 
=-\partial_i u_j^{<K}  \bigg[\big(u_i^{[P]}u_j^{[K-P,K+P]}\big)^{<K}  - u_i^{[P]}\big(u_j^{[K-P,K+P]}\big)^{<K}\bigg].$$ 
equivalent to (\ref{fund-id}) in the text. By applying the H\"{o}lder inequality and the 
$L_p$-boundedness of Fourier multipliers for polygonal partial summation, then, 
similarly as above, we obtain 
 $$ \|\Pi_K(\bu,\bu^{[P]},\bu)\|_1\leq C\|\grad\bu^{<K}\|_4 \|\bu^{[P]}\|_4 \|\bu^{[K-P,K+P]}\|_2.$$
Since \textcolor{black}{$\|\bu^{[K-P,K+P]}\|^2_2=\|\bu^{>K-P}\|^2_2-\|\bu^{>K+P}\|^2_2,$
we obtain from the first relation in (\ref{scaling})}
%2\int_{K-P}^{K+P} dk\,\,E(k),$ we obtain from (\ref{spectrum}) that 
$$ \|\bu^{[K-P,K+P]}\|^2_2=C \left[(K-P)^{-2\sigma_2}-(K+P)^{-2\sigma_2}\right]
      \leq C' P K^{-(1+2\sigma)} $$
for any $P<K/2$ and any $\sigma<\sigma_2.$ Finally, employing the estimates for 
$\sigma'<\sigma_4$
$$ \|\grad\bu^{<K}\|_{4} <C' K^{1-\sigma'},\,\,\,\,\,\|\bu^{[Q]}\|_{4}\leq C' Q^{-\sigma'} $$
from (\ref{besov-grad}) and (\ref{besov-PL}) with $p=4$ yields the final inequality. \hfill $\Box$

\renewcommand{\theequation}{A-\arabic{equation}}
\setcounter{equation}{0}  % reset counter 
\setcounter{Prop}{0}  % reset counter 
\section*{APPENDIX B}  % use *-form to suppress numbering

%\ref{Kr-transfer},\ref{Kr-GU},\ref{asympt-Kr},\ref{Kr-4third}

We provide here some details of the derivation of equations (\ref{asympt-Kr}) and
(\ref{Kr-4third}) in the text. Thus, we take $k=K-qv$ and $p=K+qu$ and evaluate
all expressions to leading order in the small parameter $\delta=q/K\ll 1.$

We begin by evaluating the interaction coefficient $\hat{B}_{kqp}$ in (\ref{Kr-transfer}).
Using the law of cosines, $k^2 = p^2 + q^2 -2pq\cos\alpha$, one obtains
\begin{eqnarray*} 
x=\cos\alpha &= & \frac{p^2+q^2-k^2}{2pq}   \cr
&=& \frac{ K^2(1+\delta \,u)^2+K^2\delta^2-K^2(1-\delta\, v)^2}{2\cdot K(1+\delta\, u)\cdot K\delta}  \cr
&=& u+v + O(\delta)
\end{eqnarray*}
In the same manner,
$$ y = \cos\beta= -(u+v)+O(\delta)\,\,\,\,{\rm and} \,\,\,\,z = \cos\gamma = 1+O(\delta^2).
$$
Thus, up to terms of relative order $O(\delta),$
$$ \hat{B}_{kqp} =4\pi^2 k^2p^2q(xy+z^3) \simeq 4\pi^2 K^4 q(1-w^2) $$ 
with $w=u+v.$

We next substitute 
$$ U(p; t|s) =U(K; t|s) + q u\frac{\partial}{\partial p} U(p; t|s) \big|_{p=K}+O(q^2)$$
$$ G(p; t|s) =G(K; t|s) + q u\frac{\partial}{\partial p} G(p; t|s) \big|_{p=K}+O(q^2)$$
and the similar expansions for $U(k;t|s),\,G(k;t|s)$ into the time-integral in (\ref{Kr-transfer}). 
Employing also the inertial-range scaling relations (\ref{Kr-GU}) and the anti-symmetry of the 
integrand in $k$ and $p$, one obtains up to terms of relative order $O(\delta):$
\begin{eqnarray} 
\lefteqn{\int_{t_0}^{t} ds \big[ G(k; t|s) U(p; t|s) -  G(p; t|s) U(k; t|s) \big] 
U(q; t|s)=q (u+v)} \nonumber\\ 
& &{}  \times \int_{t_0}^{t} ds \bigg\{   g\big(\varepsilon^{1/3} K^{2/3} (t-s) \big) 
\frac{\partial}{\partial p} \big[ \frac{C}{2\pi}\varepsilon^{2/3} p^{-11/3} r
\big(\varepsilon^{1/3} p^{2/3} (t-s) \big)\big]_{p=K}   \nonumber\\ 
& &{}-\frac{\partial}{\partial p} \big[ g\big(\varepsilon^{1/3} 
p^{2/3} (t-s) \big) \big]_{p=K} \frac{C}{2\pi}\varepsilon^{2/3} K^{-11/3} 
r\big(\varepsilon^{1/3} K^{2/3} (t-s) \big) \bigg\}
\frac{C}{2\pi}\varepsilon^{2/3} q^{-11/3}r\big(\varepsilon^{1/3} q^{2/3} (t-s) \big)  \nonumber\\ 
& &{} \,\,\,\,\,\,\,\,\,\,\,\,\,\,\,\,\,\,\,\,\,\,\,\,\,\,\,\,\,\,\,\,\,\,\,\,\,\,\,\,\,\,\,\,\,\,\,\,\,\,\,\,\,\,
          \,\,\,\,\,\,\,\,\,\,\,\,\,\,\,\,\,\,\,\,\,\,\,\,\,\,\,\,\,\,\,\,\,\,\,\,\,\,\,\,\,\,\,\,\,\,\,\,\,\,\,\,\,\,
          \,\,\,\,\,\,\,\,\,\,\,\,\,\,
          = -\varepsilon w I \frac{C^2}{(2\pi)^2}K^{-16/3} q^{-8/3}.       \nonumber\\ 
\nonumber\end{eqnarray}
To obtain the last line the integration variable was changed to $\tau = \varepsilon^{1/3} k^{2/3} (t-s)$
and the resulting integral evaluated to leading order to be $I=\frac{11}{3}I_1+\frac{2}{3}I_2,$ with
$$ I_1=\int_0^\infty g(\tau)r(\tau)\,d\tau\doteq 0.76,\,\,\,\,
     I_2 =\int_0^\infty [g'(\tau)r(\tau)-g(\tau)r'(\tau)]\, \tau\,d\tau \doteq -0.19. $$  
These latter two integrals were numerically evaluated by Kraichnan\cite{Kraichnan66}, eq.(6.2).

Finally, one obtains from (\ref{Kr-transfer}):
$$ \ol{T}(k,q,p) \simeq  -\varepsilon w(1-w^2) C^2 I K^{-4/3} q^{-5/3},$$ 
with $I\doteq 2.66$ and $C\doteq 1.77$ the Kolmogorov constant from the ALHDIA closure
(see Kraichnan\cite{Kraichnan66}, eq.(6.1)). This is exactly equation (\ref{asympt-Kr}) in the text. 

We now substitute this result into (\ref{flux-Kr3}) to obtain (\ref{Kr-4third}). The integral over $u$ 
and $v$ is easily  evaluated by the change of variables $w=u+v,\,\,t=(u-v)/2,$ giving
$$ \int_0^1 dv\int_0^{1-v} du\, w(1-w^2) =  \int_0^1 dw\int_{-w/2}^{w/2} dt \,w(1-w^2) =
      \int_0^1 dw \,w^2(1-w^2)=\frac{2}{15}. $$
      
It may be useful to discuss briefly the relationship of our heuristic derivation of  the 
$4/3$-type  scaling of $\ol{\Pi}_K(\bu,\bu^{[Q]},\bu)$ based on decorrelation effects in 
Section II.C.1 and the derivation using Kraichnan's ALHDIA closure presented in this appendix. 
The separate input  and output terms in the ALHDIA expression for $\ol{T}(k,q,p)$ each 
yield, on order of magnitude,
$$ \ol{\Pi}_K^{{\rm in/out}}(\bu,\bu^{[Q]},\bu) \simeq K^2 Q^2 \tau(K) E(K) E(Q) $$
where $\tau(K)\sim \varepsilon^{-1/3}K^{-2/3}$ is the eddy-turnover-time at wavenumber $K.$
This factor arises from the time-integration in eq.(\ref{Kr-transfer}) and represents
the effect of decorrelation in Kraichnan's closure. Using the relations that $K E(K)\simeq 
[\delta u(\ell_K)]^2$ and $K\tau(K)\simeq 1/\delta u(\ell_K),$ this result becomes
$$ \ol{\Pi}_K^{{\rm in/out}}(\bu,\bu^{[Q]},\bu) \simeq  \frac{1}{\ell_Q}
                             \delta u(\ell_K) [\delta u(\ell_Q)]^2. $$  
However, because of the near cancellation between input and output terms for 
$Q\ll K,$ the final ALHDIA result is smaller by a factor of $Q/K=\ell_K/\ell_Q,$ giving 
$$ \ol{\Pi}_K^{{\rm ALHDIA}}(\bu,\bu^{[Q]},\bu) \simeq K Q^3 \tau(K) E(K) E(Q)
                         \simeq \frac{\ell_K}{\ell_Q^2}\delta u(\ell_K) [\delta u(\ell_Q)]^2. $$
This corresponds very closely to our result and derivation in Section II.C.1. The scaling
of the partial flux $\ol{\Pi}_K(\bu,\bu^{[Q]},\bu)$ in our eq.(\ref{mean-IR}) there is due to the difference,
$$ \frac{1}{\ell_Q}\big\langle \delta u^2(\ell_Q)  \delta u(\ell_{K+Q}) \big\rangle 
- \frac{1}{\ell_Q}\big\langle \delta u^2(\ell_Q)  \delta u(\ell_{K}) \big\rangle, $$
where the two terms 
scale similarly to each other  and nearly cancel. This cancellation yields an additional 
factor of $\ell_K/\ell_Q,$ so that our final result scales as 
$$ \ol{\Pi}_K(\bu,\bu^{[Q]},\bu)\simeq \frac{\ell_K}{\ell_Q^2}
              \langle\delta u(\ell_K) [\delta u(\ell_Q)]^2\rangle. $$

\bibliographystyle{unsrt}
\makeatletter
\renewcommand\@biblabel[1]{\textsuperscript{#1}}
\makeatother

\clearpage
\bibliography{turbulence}
\clearpage

\end{document}